\documentclass[12pt]{report}
\pdfoutput=1
\usepackage{emusicthesis}
\usepackage{lscape}
\usepackage{url}
\usepackage{rotating}
\usepackage[pdftex]{graphicx}

\begin{document}
\DeclareGraphicsExtensions{.jpg,.pdf,.mps,.png}

% The title Master's Thesis 
\title{Riddim: A Rhythm Analysis and Decomposition Tool Based On
  Independent Subspace Analysis}

% This should be the author of this thesis.
\author{Iroro Fred {\d O}n{\d o}m{\d e} Orife}

% This is the name of the .bib files that holds all of your references.
% It can be a comma separated list if there is more than one file.
\reffile{references}

% These are the names of the .tex files which hold the
% respective section of the thesis.  If one of these sections
% is not included, comment out the corresponding line.
\abstract{abstract}
\acknowledgements{acknowledgements}
\dedication{dedication}

% Include the chapters.
%
  \ifnum0=\value{mychaptercount}
    \startingpages
    \setcounter{mychaptercount}{1}
  \fi
  \chapter{Introduction}

\vspace{10mm}

\begin{quote}
  {\it   
  ``Jam sessions, jitterbugs and cannibalistic rhythmic orgies are
  wooing our youth along the primrose path to Hell!''} --- The Most
  Reverend Francis J.L. Beckman in an address to the National Council
  of Catholic Women, Biloxi, Mississippi, October 25, 1938
\end{quote}

\begin{quote}
  {\it   
    ``Without music, life would be a mistake... I would only believe in a
  God that knew how to dance.''} --- Friedrich Nietzsche
\end{quote}

\vspace{7mm}
\section{Some Personal Compositional Perspectives on Rhythm}
\vspace{5mm}

My interest in understanding rhythm comes from my struggles 
to write music that incorporates rhythm in fresh 
and exciting new ways.  Frequently while listening, 
I am deluged with musical ideas for a new work or work 
already in progress.  Many times, I find it difficult 
to pick out the individual patterns or understand, in real 
time, how different components of a percussion mix interrelate.  
For the experienced listener and musician this might be a 
routine task, but I would argue that such transparency is 
less common in electronic or dance styles that play with the 
human perceptual limits (Aphex Twin, Squarepusher, etc) especially 
where the recording is the only source of information about the 
underlying musical and rhythmic structure.

On one hand, repeated listenings and intensive study would facilitate 
a better understanding; however, for the compositional 
novice, a tool that embodied the skills of ear-training and expert 
knowledge on rhythm perception that could uncover the hidden 
structure and more complex inter-relationships between various 
elements of a mix, would be invaluable. 

Another compositional use of such a tool would be to aid mapping out 
new rhythmic configurations in a more orderly fashion.  Rather than 
employing music theoretic heuristics with ad-hoc trial and error 
experiments, new patterns could be systematically found by traversing 
{\it Rhythmic Spaces}, (defined below). The problem with trial and error 
is that it is time consuming. Managing the complexity of the
interrelationships between patterns and where they reside in the 
rhythmic space becomes unwieldy as the said patterns accumulate. With a 
computational tool, such an effort would be markedly simpler 
especially if rhythmic features are modeled parametrically. 
Several features that might be interesting to explore are 
onset timing intervals, accentuation of onsets, overall tempo 
and timing and grouping hierarchies between elements of a mix. 
A ``rhythmic space traversal'' would take a rhythmic pattern 
with certain of the above features and hold all but one constant 
while creating new patterns that varied the last feature in some way.

\begin{figure}[thp]
  \begin{center}
    \resizebox{3.5in}{!}{\includegraphics{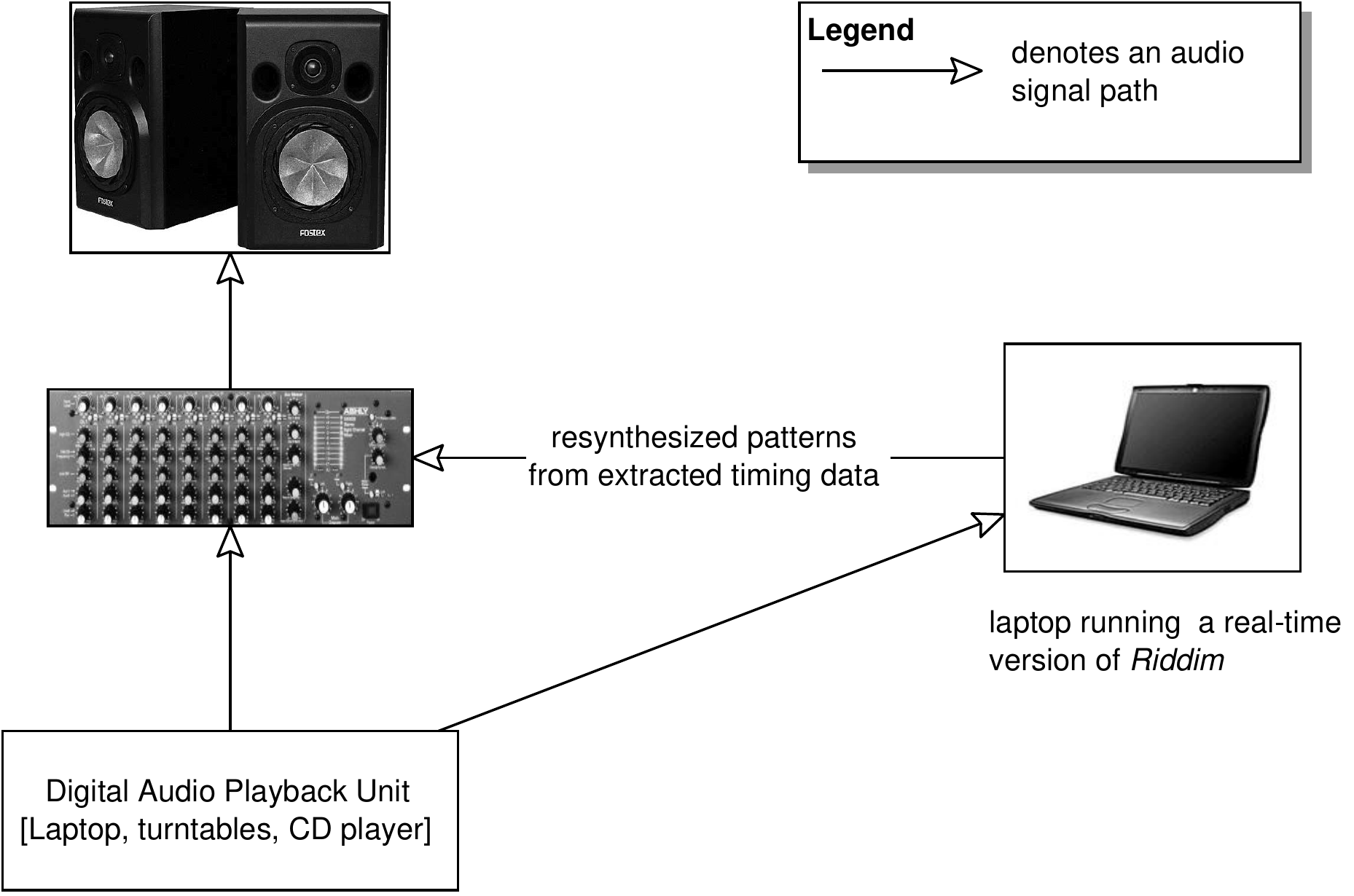}}
    \caption{Live performance with Riddim}
    \label{A real-time possibility of Riddim} 
  \end{center}
\end{figure}

\vspace{7mm}
\section{Towards a General Perceptual, Transcriptive Computational Tool}
\vspace{3mm}

A perceptual {\sl and} transcriptive tool is most useful to meet the 
compositional objectives described above.  It needs to be
transcriptive because resynthesis can only take place from an abstract
representation. It also needs to be perceptual because no score or
advanced knowledge about the structure of the music may be available.
In this case, any structure the listener perceives is inferred
via direct perception.  This is especially true of many styles,
especially ones practiced solely in oral traditions. 
Is everything that is needed for a perceptual transcriptive
representation of rhythm present in the recorded audio?  The
information {\sl must} be contained in the structure of the audio
waveform because it is this same audio that is processed 
and understood by the brain. 

In the past, researchers have tended to focus on a very specific area 
of rhythm, be that tempo, beat-induction, metrical interpretation or 
rhythmic pattern recognition. Only rarely did they ever integrate 
their work in a framework concerned with the greater experience of 
rhythm perception.  What is attempted here is the development of a 
tool that explores a more integrative approach. Integrative is used 
in the sense that the streams of timing sequences extracted from an audio 
file can be incorporated in existing interpretive models of rhythm, 
that together can approach a computational correlate of a ``rhythmic percept''.

\vspace{7mm}
\section{Thesis Structure}
\vspace{3mm}

This thesis is structured in the following way. First, I will review the
most salient literature on rhythm, by going through its various dimensions, 
discussing the history and varying opinions. Next I shall discuss my 
approach to developing {\it Riddim}, a rhythm analysis tool designed 
to attempt to meet some of the compositional needs described above.  
This will include some high level algorithmic descriptions of my approach, 
a brief description of the development platform and how the various 
components operate.  Next, I will explain the workings of each of the 
algorithms and illustrate why they are important and how they are 
used, providing detailed diagrams of how data flows through each module.

Finally, I shall show my results which are in-depth rhythmic analyses 
of a variety of musical recordings using {\it Riddim}. {\it Riddim} will
be distributed as a stand alone software application from which commands to
perform a variety of rhythmic analyses can be executed. Results can then
be viewed and saved as MIDI files or rendered as digital audio for use
in subsequent creative settings.

  \ifnum0=\value{mychaptercount}
    \startingpages
    \setcounter{mychaptercount}{1}
  \fi
  \chapter{Rhythm}
\vspace{10mm}

\begin{quote}
  {\it   
  ``Compositional activity proves meaningless in the
  absence of a corresponding bodily gestalt.''} --- Ellington's Law 
by Warren Burt 
\end{quote}

\begin{quote}
  {\it ``Jazz without the beat, most musicians know, is a telephone yanked
  from the wall, it just can't communicate.''} --- Leonard Feather,
British author and jazz enthusiast.
\end{quote}

\vspace{7mm}
\section{Understanding Rhythm Through Modeling}
\vspace{5mm}

A fundamental assumption in this work is that representation and modeling
are ways to understand complex real-world processes, in particular, 
the activity of human perception of rhythm. Theories can then be
formulated and verified using empirical data and models, which in 
turn can be updated to reflect new knowledge.  Thus, modeling the cognitive 
processes associated with rhythm and music in general, can yield better
understanding of the functions underlying specific behavioral
responses to music and rhythm.  This knowledge is potentially useful 
to composers, performers, designers of codec\footnote{The term codec 
stands for ``compression/decompression''. A codec is an algorithm, or
specialized computer program that does exactly that. Popular audio
codecs include MPEG 1 Layer 3 (mp3) and Sony minidisc.} 
algorithms, interactive music and media systems \cite{Desain:92}.
These models embody functional relationships between physical
properties of sound and resulting human behavior and sensations.
Thus it is important to make distinctions between the physical
properties of sound, called {\it phenomenal}, that can be measured and their
perceptual correlates.  
%
% looks good as is
%\begin{center}
%
\begin{table}
\begin{tabular}{|l|l|} % all cell contents aligned center
 \hline 
 & \\
 Physical properties of sound  & Perceptual correlates \\ \hline \hline
 & \\
 Frequency & Pitch   \\ \hline
 & \\
 Signal Intensity & Loudness   \\ \hline
 & \\
 Differences in sounds with identical loudness and pitch & Timbre \\ \hline
\end{tabular}
\caption {Properties of sound and their perceptual correlates}
\end{table}
%\end{center}

\vspace{7mm}
\section{Time}
\vspace{3mm}

\subsection{Temporal Representations in Music}

Music is time-based art form that always occurs in some cultural
context. Accordingly, the representation of musical time is 
culturally dependent. Different cultures have time represented 
differently in their cognitive structures, for example, in some 
contexts, time is seen as a circular structure, rather than a
continuum \cite{Gardenfors:2000}. Musical enculturation is thus a 
combination of perception and the organisation of sounds based on 
internalised cultural templates \cite[p. 5]{Smith:99}. Different areas
of research also have different perspectives and interests in
representing time. For example, while musicologists might be
interested in representations that facilitate notation and transcription,
composers and performers might be more interested in representations that 
are designed to work in process-oriented real-time systems \cite{Honing:93}.

There are several different views of temporal representation in
music. These include {\it tacit}, {\it implicit} and {\it explicit}
time. Tacit time frameworks do not represent temporal evolution; there is 
only an idea of ``now''. Implicit time structures employ no definitive 
time relations since time is represented as an absolute, e.g. note
lists, while explicit time structures represent time as relations that
can be compounded to form higher level notions of time. Other
questions that come up in the time modeling process deal with absolute
versus relative time bases, discrete versus continuous representations
and point-based versus intervallic time primitives. Point-based
implies that events are duration-less and occur serially while
intervallic refers to differences between events that form the basis
for meaningful relations (intervals that occur before, after, during) \cite{Honing:93}.

\begin{table}
\begin{tabular}{|l|l|}\hline % | is used for a vertical line
& \\
Low Level: Expression  & {\it Perceptually represented} as departures \\
                       & from canonical proportional values; poorly \\
                       & quantified, and experienced as expressive \\ 
                       & rather than durational effects. \\ 
                       &  \\
                       & {\it In Performance}, represented as programmed\\
                       & variations in the rate of the clock controlling\\
                       & beat durations (see level 2), and as modifications\\
                       & of the procedures specifying individual notes\\
                       & \\ \hline
                       & \\

Middle Level: Rhythm   & {\it Perceptually represented} as collection \\
and meter              & of grouped durational equalities and \\
                       & inequalities organized around a metrical \\
                       & framework (when the music has a meter) \\
                       &  \\
                       & {\it In Performance}, represented as a \\
                       & collection of untimed procedures, organized \\
                       & metrical markers which are directly timed by \\
                       & a programmable clock. \\
                       & \\ \hline
                       &  \\

High Level: Form       & {\it Perceptually represented} as a structure\\
                       & of hierarchical relations, constructed by means\\
                       & of memory processes and perception, and \\
                       & distinguished from level 2 structures by \\
                       & exceeding the length of the perceptual present \\
                       &  \\
                       & {\it In Performance}, represented as a hierarchical \\
                       & memory structure that forms the highest levels\\
                       & of a motor program \\
                       & \\ \hline
\end{tabular}
\caption {A summary of structural levels in musical time based on
  meter and their cognitive/perceptual properties \cite[p. 233]{Clarke:87}.}
\end{table}

\vspace{5mm}
\subsection{Causal versus Non-Causal Analysis}
A further point regarding time arises when an interpretive process must
examine some data structured in time. Some theories propose digesting
whole chunks of data as a unit while others choose to process
the data sequentially in time or causally. Opponents of the former
process, claim that ``theories that behave symmetric with respect to 
time have to be wrong on that basis alone. ... [those] theories 
tend to model the perceiver more like a musicologist studying the
score, instead of a first time listener'' \cite{Desain:92}. On the
other hand, proponents claim that the non-causality of such models
embodies a certain amount of enculturated knowledge that is essential to
any interpretation, citing that performers have spent time with their
music and have definite intentions before a show \cite[p. 55]{Smith:99}. 

Analytically, causal methods can be seen as a subset of the larger
non-causal methods, even though the tools may not be related.
In any case, because humans infer musical structure based on
both universal ``hard-wired'' low level processing of sensory data
(bottom-up processing) {\it and} culture specific knowledge,
expectations and predictions (top-down processing)
\cite[p. 27]{Todd:94}, robust models of rhythm perception must account 
for both bottom-up and top-down processing and cross-cultural variances. 
This makes room for multiple interpretations in which both causal and
non-causal methods are useful.

\begin{table}
\begin{tabular}{|l|l|}\hline % | is used for a vertical line
& \\
0 to 2-5 msecs apart   & Beats are perceived as simultaneous, \\
                       & indistinguishable (even with different \\
                       & loudness, but same duration) as a single\\
                       & event. \\
                       & \\ \hline
                       & \\

2-5 to 40 msecs apart  & Beats are distinguishable, but no order \\
                       & relation can be indicated. \\
                       & \\ \hline
                       &  \\

30 to 50 msecs apart   & Beats above this can produce an order \\
                       & relation. \\
                       & \\ \hline
\end{tabular}
\caption {A table summarising Leigh Smith's review of P{\"o}ppel's
 taxonomy of elementary time experiences \cite[p. 14]{Smith:99} }
\end{table}

\vspace{5mm}
\subsection{Specious Present and Auditory Stores}

Another aspect of time deals with the interval in which
events are processed -- the {\it specious} present. Later called the 
psychological, perceptual or subjective present, it is related to the 
duration within which one experiences a sequence of events as
simultaneously present in consciousness \cite[p. 97]{Gabrielsson:89}\cite{Fraisse:63}.
Parncutt defines this perceptual present as a ``continuous time
interval comprising all real-time percepts and sensations
simultaneously available to attention, perception and cognitive 
processing'' \cite[p. 451]{Parncutt:94}\cite[p. 12]{Smith:99}.

Enabling this present is {\it echoic} memory, a kind of auditory
sensory memory possessing a high degree of structure and organization \cite[p. 451]{Parncutt:94}. 
Echoic memory which is acoustic should be differentiated from {\it iconic} memory which 
is visual. There have been many different estimates in the literature 
on the size of echoic memory and the ``integrating buffers'', most of 
them limiting the echoic memory to less than 500 msec and the
subjective present to 4 seconds, with experimental results ranging 
between 2 and 10 seconds \cite[p. 12-3]{Smith:99}
\cite[p. 451]{Parncutt:94}. Parncutt stresses that because the
``present is so highly adaptive, no fixed parameter values can be
expected to describe it adequately'' \cite[p. 451-2]{Parncutt:94}. 
Furthermore, the present also depends on the rate events occur and 
their complexity and structure \cite[p. 452]{Parncutt:94}.

\vspace{5mm}
\subsection{Synchronisation}

Synchronisation is another aspect of rhythm that emerges from a notion of
time. Mari Riess Jones in an article entitled ``Time, Our Lost
Dimension: Toward a New Theory of Perception, Attention and
Memory'' states that the ``human system in general and the perceptual
system in particular depend upon the properties of endogenous
rhythmic processes in the nervous system'' \cite{Jones:76}. Additionally, 
many studies have suggested the presence of internal clocks and have
noted the strong connection between physical actions and perception 
\cite[p. 17]{Smith:99}. In reaction to a sound stream, humans are able
to forecast and coordinate motor action to coincide with future events
from the sound stream. Smith claims that regularity is less important 
to the act of synchronisation than the listener's expectations and ability to
predict, since accelerating and decelerating rhythms can be
timed \cite{Smith:99}. Jones suggests that parts of the synchronisation process are
based on temporal expectancy.  Thus relations between certain 
time differences (time deltas) will hold for subsequent ones. This allows
the listener to extrapolate time trajectories that predict when events
will occur \cite{Jones:76}. 

\vspace{7mm}
\section{Definitions}
\vspace{3mm}

How should we define rhythm?  Smith reports a variety of responses from the literature
\cite[p. 9-10]{Smith:99}. Eric Clarke used the term broadly to apply
to ``regular, periodic features of the temporal structure of music and
to aperiodic features'' \cite{Clarke:85structure}. Dowling states that
rhythm is a temporally extended pattern of durational and accentual 
relationships, while Parncutt claims that musical rhythm is an
acoustic sequence evoking a sensation of pulse
\cite[p. 451]{Parncutt:94}.  Scheirer cites Handel's claims that
``the experience of rhythm involves movement, regularity, grouping,
and yet accentuation and differentiation'' \cite{Scheirer:98tempo}.

According to psychological studies done by Gabrielsson on human 
``subjects'', rhythm like timbre is a multi-dimensional quality. 
From the empirical studies, Gabrielsson points out the existence of at 
least fifteen dimensions which ``lend themselves to grouping into 
three categories'' related to structural, motional and emotional
aspects of rhythm. Refer to Table 4.

\begin{table}
\begin{tabular}{|l|l|}\hline % | is used for a vertical line
& \\
Structural Aspects     & Meter, position and strength of accents, \\
                       & type and prominence of basic pattern, number \\
                       & of different kinds of subdivisions within \\
                       & beats, uniformity versus variation and  \\
                       & simplicity versus complexity \\
                       & \\ \hline
                       & \\

Motional Aspects       & Tempo and overall rapidity, however motional \\
                       & characters exist such as walking, dancing,  \\
                       & jumping, rocking, swinging, graceful and driving \\
                       & forward \\
                       & \\ \hline
                       &  \\

Emotional Aspects      & Vitality versus dullness, excitement versus \\
                       & calmness, rigidity versus flexibility and  \\
                       & solemnity versus playfulness \\
                       & \\ \hline
\end{tabular}
\caption {Gabrielsson's dimensional categories for rhythm \cite[p. 107]{Gabrielsson:89}}
\end{table}

There are problems in modeling the ``dimensions'' mentioned above
because the interpretive ambiguity of the dimensions and characters 
do not lend easily to symbolic or quantitative representation.

\vspace{7mm}
\section{Pulse}
\vspace{3mm}

Parncutt's definition of a musical rhythm is an acoustic
sequence evoking a sensation of pulse.  How should we define pulse?  
If a car drives by loudly broadcasting music, in the several seconds within
which the music is in earshot, the pulse would be the subjective
evaluation of the feel or impression of movement inferred from the
music. If that music happened to be tabla drumming, Olatunji, Faithless or
Subotnick, each would leave a different but firm impression in the mind
of the listener.  
The percept of pulse is confined to the time interval called the
subjective present \cite[p. 451]{Parncutt:94}. While {\it rhythm}
deals with grouping and time hierarchies and {\it tempo} involves the 
rate at which rhythmically significant events occur, pulse corresponds
to a ``sense of equally spaced phenomenal impulses''
\cite{Scheirer:98tempo}. For many kinds of music, the pulse is the
``foot-tapping'' beat.

Pulse has also gone by other names like {\it tactus}, defined by Lerdahl and
Jackendoff to describe the most salient hierarchical level at which
the listener will tap their foot in accompaniment to a rhythm \cite[p. 31]{Smith:99}.
Parncutt defines it as a trivial sense of expectancy. Once a pulse is
established, subsequent events are ``expected'' with some time
trajectory. This is of course related to our discussion above of
synchronisation, where motor actions move along time trajectories
inferred from this expectation \cite[p. 453]{Parncutt:94}.

\vspace{5mm}
\subsection{Pulse Outside A Metrical Context}

In non-metric musics, the pulse may not necessarily correspond to an
isochronous period.  Researchers agree that the idea of a basic pulse
in such music is ``usually misplaced'' \cite[p. 190]{Magill:97}
Magill reports that 
\begin{quote}
``West African drum ensemble music takes a multilayered approach to
rhythm, with all the parts being related to a fundamental {\it time
line}, often played on a bell... Bell patterns can nearly always be
subdivided into a number of regular pulses (usually 8, 12, or
16). Time lines or bell patterns betray asymmetric construction
and sound syncopated to Western ears'' \cite[p. 190]{Magill:97}.
\end{quote}

For example, in the A{\it n}lo drumming style from southern Ghana, 
Pantaleoni points out that it is ``the bell that regulates the play, 
and not the steady pulse some player or observer might feel, because 
the bell can put each pattern into its proper places, while a simple
pulse can only regulate the speed with which the pattern is played''
\cite[p. 60]{Pantaleoni:72}. Magill states that in many West African 
drumming styles while some instruments reinforce an underlying fast pulse, 
others might highlight portions of a bell pattern while others provide
a complex overlay of rhythms and meters \cite[p. 190]{Magill:97}. In
any case, while regular pulses are quite common, the traditional 
cultural view is that they are of peripheral importance compared 
to the asymmetric time line. The definition of pulse needs to be 
revised to be a periodic structuring or synchronization framework, 
since in the kinds of music discussed above, asymmetric time lines 
and isochronous pulses are both constructs within which higher levels 
of rhythmic time are structured. 

\vspace{5mm}
\subsubsection{ The Fastest Pulse }

Attempts to understand non-metric music have also emerged in the form of
concepts like the ``fastest pulse''. The fastest pulse is defined as 
the ``beat with the shortest duration in the music considered''
\cite[p. 33]{Smith:99} Smith cites Koetting's warning that the fastest
pulse does not seem to describe how African timing is perceived
\cite[p. 33]{Smith:99}.  Pantaleoni comments \begin{quote} ``I have
not found it to be a concept familiar to those African drummer with 
whom I have worked, nor does it seem to be an easy concept for them to
use once it has been explained. Perhaps it is simply difficult for a 
performer to think as an analyst, but in the absence of some kind of 
positive support from the musicians themselves it cannot be assumed 
that a convenient analytical tool corresponds to the basic pulse or 
principle of timing which actually functions in the playing of the 
music'' \cite[p. 58]{Pantaleoni:72}. \end{quote}
While clearly limited from a compositional and improvisational point 
of view, the idea of the fastest pulse might still be useful in
analysis. As will be seen in the next chapter, analyses that determine
the fastest pulse, per extracted stream, are a useful interpretive tool.

\vspace{5mm}
\subsection{Beat Induction}

Beat induction is a synchronization process with a phenomenal pulse.  
There have been many successful attempts to measure the pulse 
computationally in Western classical and popular music, using that 
information to drive a foot-tapping process.  
Eric Scheirer used filterbanks and parallel comb filters to extract
amplitude envelopes and infer the beat from arbitrarily complex 
musical signals \cite{Scheirer:98tempo}.  Povel and Essens introduced 
the idea of ``internal clocks'' which are selected based on a set of 
inter-onset interval inputs. Desain and Honing have several computational beat-tracking 
methods to their name. Given an onset stream, their models 
return the beat of the stream. Different from the 
previous methods, Desain and Honing use a causal process model which takes the
onset inputs sequentially and updates an expectancy measure.  The
beat is then inferred from regions of high expectancy. Large and Kolen's
beat-tracking model is based on non-linear oscillators. From a stream
of onsets, a ``gradient-descent method'' is employed to continually
update the period and phase of an oscillator, which represents the
pulse \cite{Large:94}\cite{Scheirer:98tempo}. Goto presented a system 
which ``combines both low-level ``bottom-up'' signal processing and 
high-level pattern matching and ``agent-based'' representation to 
beat-track and do simple rhythmic grouping for popular music''
\cite{Scheirer:98tempo}\cite{Goto:98}. Leigh Smith has also explored a
multiscale representation of musical rhythm. Onset streams or pulse
trains are first decomposed using a Morlet wavelet basis. The
``foot-tapping'' pulse of a section of music is then extracted by 
correlating frequency modulation ridges extracted from the wavelet
decomposition using stationary phase, modulus maxima, dilation scale 
derivatives and local phase congruency. Refer to \cite{Smith:99} for 
definitions and implementation details.

\vspace{5mm}
\subsection{Tempo}
Tempo is defined here as the rate at which rhythmically significant
events occur. In music with an isochronous pulse, tempo is the rate of
the tactus. In much Western music and popular music, this is the ``foot
tapping'' rate. In non-metrical settings, there may not be any explicit
overall tempo. In this case, there may be multiple rhythmic layers
each with its own individual rate. Tempo is however best used in a 
metrical setting. Commercial software systems express tempo in terms 
of beats per minute (BPM), while numerous scores have subjective tempo 
indications like {\it Larghissimo} or {\it Prestissimo} or more
concrete markings like quarter note equals sixty.

Within the above framework how do variations in tempo affect the
interpretation of rhythmic structure?  Clarke reports that slow tempi
aid the cognitive process of segmentation at structural boundary
points while at higher tempi, music tends to be grouped into fewer
units. This is claimed to occur because of the limits of the subjective
present. In other words, at faster tempi, more elements are packed into
the subjective present; thus given fixed maximum neural processing
rates, music is subdivided into larger groups in accordance with 
its structural properties \cite[p. 36]{Smith:99}.

\vspace{7mm}
\section{Grouping}
\vspace{3mm}

Experiencing rhythm occurs as a whole sequence or pattern, not as
individual events. Within the subjective present, how a sequence 
is formed by segregating a stream of events is called grouping. This
can take the form of phrases or motives. In many cases, the number of 
events grouped together depends on the presentation
rate, so the faster the rate, the more members are included in the
group. This extends up to perceptual limits.  Some rhythmic 
groups are presented so quickly that they're not perceived as
individual rhythms or events but are lumped together to form a pitch percept. 
Within a sequence of events, certain ones with objective accents or 
differences suggest or confirm the presence of a perceptual group
boundary.  Quantities responsible for the percept of a group 
boundary include perceptual limits on memory and attention and 
cognitive representations.

Smith points out two principal grouping principles: The {\it run} 
principle and the {\it gap} principle.  The run principle proposes that 
the longest run of similar events will begin a rhythmic pattern. 
Listeners tend to group sounds of the same intensity together. Runs of 
different intensity will be organized so that the longest runs are
placed at the beginning or end of the pattern, never in the middle. 
The gap principle refers to boundaries that are formed between 
dissimilar elements. Rests partition elements while elements close 
in time tend to be grouped \cite[p. 28]{Smith:99}. These principles
are based on Gestalt principles of perceptual organization and
Bregman's seminal work on auditory scene analysis
\cite{Bregman:90}. Additionally, higher level grouping structures are 
bounded by repetitions. A single change in an established pattern 
can affect the entire grouping percept \cite[p. 28]{Smith:99}.

Parncutt defines two independent kinds of temporal grouping, 
{\it serial} and {\it periodic}. Serial grouping (related to the gap
principle) is dependent on the serial proximity of adjacent events in
time, pitch and timbre. ``According to Lerdahl and Jackendoff (1983), 
serial grouping in music includes ``motives, themes, phrases, periods,
theme-groups, sections, and the piece itself'' \cite{Lerdahl:83}. 
Periodic grouping on other hand, depends on the relative times and 
perceptual properties of {\it nonadjacent} events \cite[p. 412]{Parncutt:94}.

\vspace{5mm}
\subsection{Meter}

Meter can be seen as a form of periodic grouping. Parncutt states that it
groups events into equivalence classes, such as all {\it n}th beats of
a bar \cite[p. 412]{Parncutt:94}. He further classifies periodic
grouping as being an aggregate of two stages, {\it pulse sensation} and
{\it perceived meter}. Interestingly enough, Halsband conversely
states that grouping ``plays a major role in the perception of all metrically
organized music'' \cite[p. 266]{Halsband:94}.  

While grouping is a psychological term, meter is a product of music
theory and associated forms of notation. Repetition in meter, as in 
other groupings, plays an integral role in the determination 
of a metrical percept. However, the perceived meter in a performance 
may not necessarily correspond to the notated meter. In many of these 
cases, some initial metrical orientation influences the interpretation
of subsequent sequences.  

There are rarely explicit rules to outline an interpretation; in 
improvisational styles, performances ``unbounded by a strict
metric frame are not free in the sense of being unrhythmic. Rather, 
they are driven by rhythmic goals that are elastic'' \cite[p. 158]{Berliner:94}.

Handel showed empirically that difficult rhythms with atypical
metrical accenting were perceived in terms of elemental grouping
schemes, rather than as a meter with a timing interval \cite[p. 31]{Smith:99}.
Thus, a metrical percept is a regular alternation of strong and weak
beats. What then qualifies an event as particularly strong or weak?

\begin{table}
\begin{tabular}{|l|l|}\hline % | is used for a vertical line
& \\
Well-Formedness Rule 1 & Every attack point must be associated with \\
                       & a beat at the smallest level metrical level \\
                       & present at that point in the piece. \\
                       & \\ \hline
                       & \\

Well-Formedness Rule 2 & Every beat at a given level must also be a \\
                       & beat at all smaller levels present at that \\
                       & point in the piece. \\
                       & \\ \hline
                       & \\

Well-Formedness Rule 3 & At each metrical level, strong beats are \\ 
                       & spaced either two or three beats apart. \\
                       & \\ \hline
                       &  \\

Well-Formedness Rule 4 & The tactus and immediately larger metrical \\
                       & levels must consist of beats equally spaced \\
                       & throughout the piece. At subtactus metrical  \\ 
                       & levels, weak beats must be equally spaced \\
                       & between surrounding strong beats. \\
                       & \\ \hline
\end{tabular}
\caption {A summary of Lerdahl and Jackendoff's Metrical Well-Formedness
  Rules from their ``Generative Theory of Tonal Music'' \cite{Lerdahl:83}.}
\end{table}

\vspace{5mm}
\subsection{Accentuation}

Accents are differences in adjacent events that separate them and suggest 
particular grouping configurations. Smith cites Lerhdahl and
Jackendoff as distinguishing three kinds of accents, {\it phenomenal},
{\it metrical} and {\it structural} according to their effect on 
groups \cite[p. 19]{Smith:99}. Phenomenal accents are said
to exist superficially, stressing a single moment and enabling
syncopations. Metrical accents occur when the emphasised beat is
part of a repeating metrical pattern. Structural accents occur at
points hierarchically higher than meter. For example, structural accents 
could delimit phrases or sections.  Furthermore, a ``phenomenal accent
functions as a perceptual input to metrical accent'' \cite{Lerdahl:83}.

Accents occur when there are changes in duration, inter-onset 
interval (IOI), pitch, articulation, timbre or combinations thereof. 
Similarly changes in phrase length, time expectancies, event density 
(flams, fills, etc) or multiple event synchronisation can cause
accentuation \cite[p. 19]{Smith:99} \cite[p. 426]{Parncutt:94}. 

Articulation is one way to signify an accent. Berliner remarks that in
the jazz idiom, ghosting ``on-beat pitches in an eight-note sequence
creates de facto accents on every off-beat, increasing rhythmic
tension. Reversing the procedure relieves rhythmic
tension... Similarly the varied application of hard, soft and ghosted
attacks can shift accents within a repeated eight-note triplet, changing
its perceived rhythmic configuration and its relative syncopated
quality'' \cite[p. 156]{Berliner:94}.

\begin{figure}[thp]
  \begin{center}
    \resizebox{5in}{!}{\includegraphics{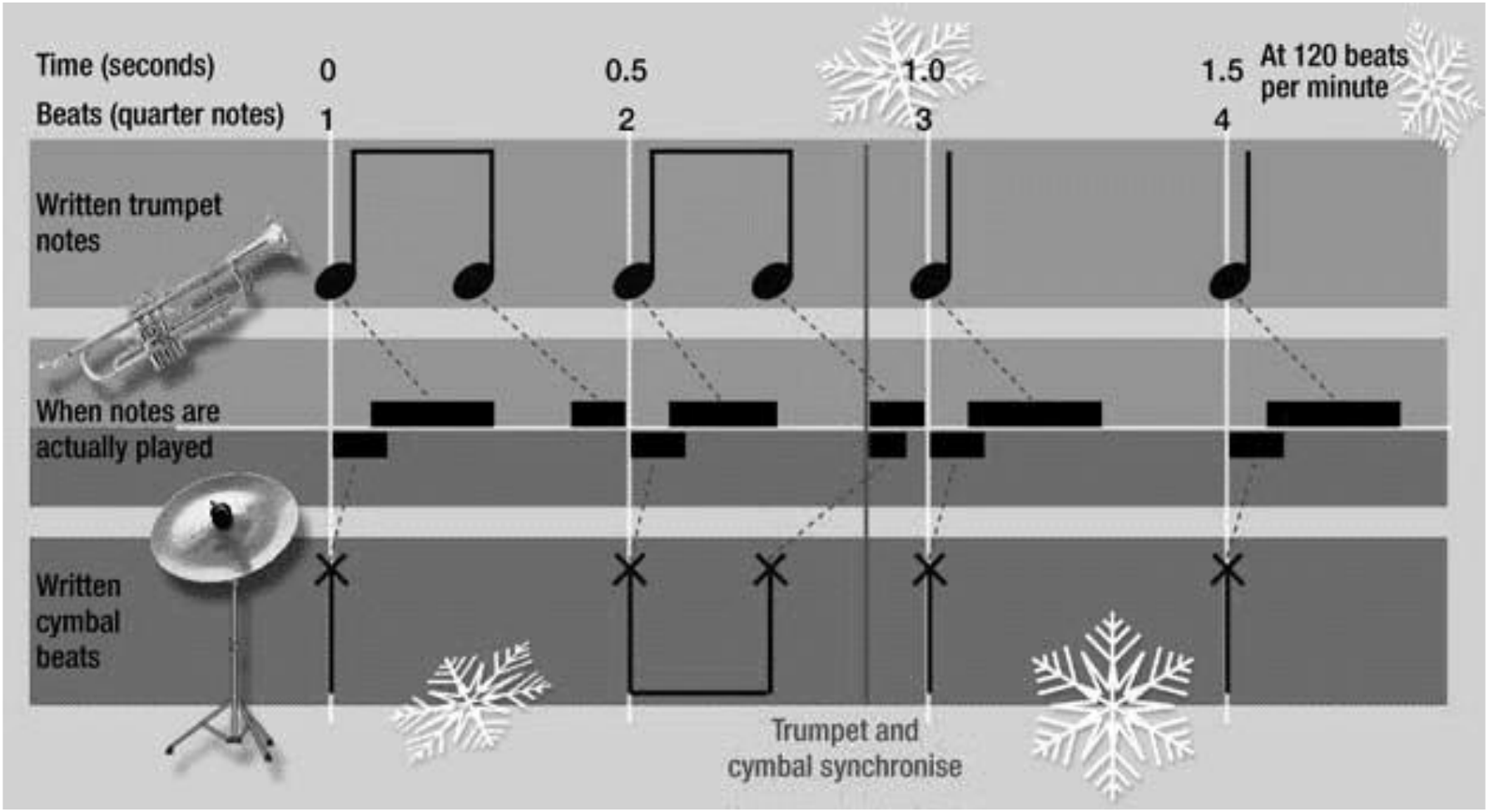}}
    \caption{``For a swinging performance the first of each pair of
      eight notes is played longer than the second. The melody line
      also hangs behind the cymbal beat, except for the occasional
      off-beat synchronisation, which keeps the band together'' \cite{Hamer:2000}}
    \label{allthatJazz}
  \end{center}
\end{figure}

Accentuation is not always applicable to specifying grouping structure
and its use varies with the musical context. For instance, using the 
A{\it n}lo dance drumming example, Pantaleoni reports that ``variation
in loudness (phenomenal accents) is indeed an important part of the 
character of a rhythmic line in Atsi{\~ a}, but no evidence indicates 
that its contribution is more than melodic... Dynamic stress would
seem to be individual, decorative, incidental, and often illusory, 
hardly a likely clue to rhythmic organisation'' \cite[pp. 54]{Pantaleoni:72}.

\vspace{7mm}
\section{Expressive Timing}
\vspace{3mm}

Many authors have explored what is known in rhythmic performances as
expressive timing. Discrepancies can always be found in the timing
dictated by a score and that which occurs in a performance. Barring
mistakes, which really are not plausible with highly trained musicians
familiar with the music, these deviations are ``related to the
structural properties of the music and to the ways the performers
organize'' these properties \cite{Clarke:85structure}. In an analysis 
of Erik Satie's ``Gnossienne No. 5'', Clarke reports that 
\begin{quote} ``the tempo marking is followed by the instruction ``souple et
expressif''. This is relevant to subsequent analysis, since it
suggests that the performance of the piece should avoid metronomic
tendencies in the left hand, and may encourage a rhythmic flexibility
in the right hand that conventional notation cannot convey'' \cite[p. 300]{Clarke:85}.
\end{quote}

Research has thus focused on determining exactly what deviations from 
the score contribute to a subjective feeling of expression that is 
not ``metronomic''. First some definitions: expressive timing refers 
to deviations in notated times that suggest particular structural 
interpretations or give the temporal development of the music a 
particular feel or sense of movement. Smith cites certain kinds
of expressive timing as having evolved from a performance practice 
known as {\it rubato}, literally ``robbed time''
\cite[pp. 37]{Smith:99}. In computer music, terms like ``micro-tempo''
are used to describe this expressive phenomenon of local tempo
changing from event to event. In all cases,
these expressive timing transformations are used by performers either 
in response to structural features in the score or in attempt to
explore new interpretive configurations  or impose a particular
structure on structurally indeterminate material \cite{Clarke:85structure}.

\begin{figure}[thp]
  \begin{center}
    \resizebox{4.5in}{!}{\includegraphics{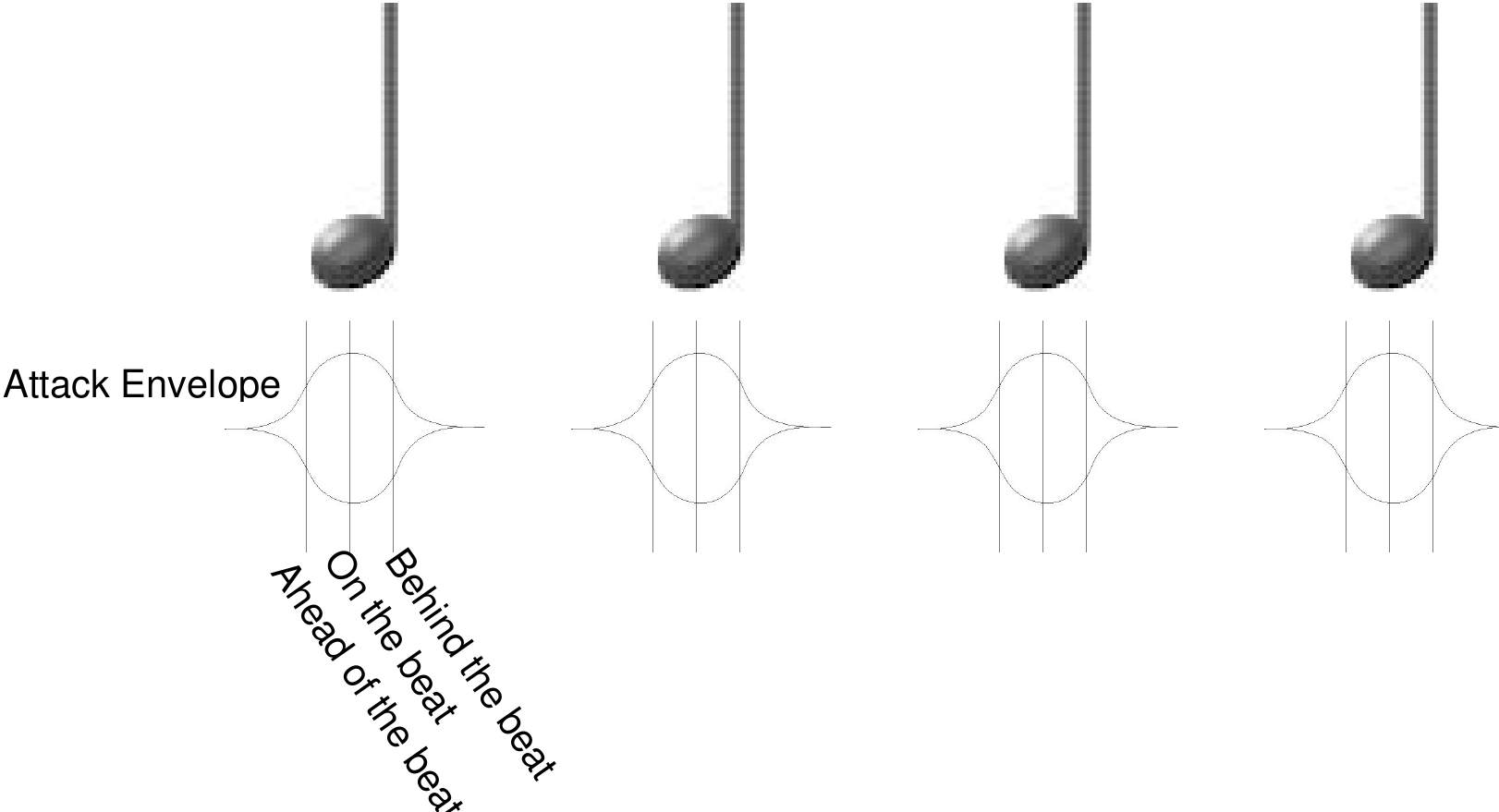}}
    \caption{``Imagining the beat as an ``elliptical figure'', the
      drummer or bass player can play either ``ahead of the beat''
      (that is, on the front part of the elliptical figure), ``behind
      the beat'', (that is on the very end of the elliptical figure or
      in varying degrees toward the center of the figure or ``on the
      beat'' (that is, the center of the figure)'' \cite[p. 151]{Berliner:94}.}
    \label{on the beat}
  \end{center}
\end{figure}

In the jazz tradition, Berliner says that musicians ``talk about 
playing on three different parts of the beat without making any
difference in the overall tempo''. On a similar note, Fred Hersch
claims that \begin{quote} ``there should be ten, fifteen different
  kinds of time. There's a kind of time that has an edge on it for a
  while and lays back for a while. Sometimes it rolls over the bar,
  and sometimes it sits more on the beats.  That's what makes it
  interesting. You can set a metronome here and by playing with an
  edge or playing behind it or right in the center you can get all
  kinds of different feelings. That's what makes it come alive. People
  are human, and rhythmic energy has an ebb and flow'' \cite[p. 151]{Berliner:94}. \end{quote}

\vspace{5mm}
\subsection{Tempo Curves}

Tempo curves are one way that timing deviations are represented in
contemporary music software applications for notated music, offering 
a measure of ``beat-by-beat time deviation from a canonical metrical grid'' \cite[p. 39]{Smith:99}.
This quasi-instantaneous tempo or {\it Local tempo} is defined
computationally as the event to event ratio of score time interval
to performance time interval. $$ Local\ tempo = Global \ tempo \times {Score \ time \ interval
  \over Performance \ time \ interval}$$ Tempo curves are thus 
constructed by connecting points with straight line segments between local
tempo measures of each performed note in the score \cite{Smith:99}\cite{Desain:91}.

\vspace{5mm}
\subsection{Expression and Structure}

Desain and Honing warn that while they are useful to study expressive
timing, tempo curves can be is a ``dangerous notion''. This is because
it encourages ``its users into the false impression that it has a musical and
psychological reality. There is no abstract tempo curve in the music
nor is there a mental tempo curve in the head of the performer or
listener'' \cite{Desain:91}. This is because tempo curves by themselves
contain no structural information. Thus global transformations, like 
scaling the overall tempo curve, will change timing relationships 
between notes that are ornamental and {\it invariant} with tempo. 
Furthermore, attempts to impose a tempo curve from the performance of 
one piece onto the score of another shows that tempo curves convey an 
erroneous notion that time is independent of the note events that mark it.
This clearly ignores the fact that the expression captured by the
tempo curve can only function with respect to the original
music performance from which it was derived \cite{Smith:99}\cite{Desain:91}\cite{Desain:91towards}. 

For a more in-depth discussion on expressive temporal transformations,
refer to Desain and Honing's paper called ``Towards a Calculus for 
Expressive Timing in Music'' \cite{Desain:91towards}.

\vspace{7mm}
\section{A West African Concept of Rhythmic Time}
\vspace{3mm}

\subsection{Misunderstandings...}

The relative lack of serious research on rhythm in sub-saharan 
African music is frankly baffling, given traditions so rich in 
rhythmic textures and structures. The {\sl Journal de la Discotheque 
Internationale de Musique Africaine} and the rare article in {\sl
Music Perception} are the only academic journals where such music 
is even discussed. 

On the one hand, this paucity of work is understandable as most of the 
research in cognitive sciences is being done in Western universities 
with funding from local or national organizations; this is compounded 
by the fact that even the relatively simple rhythmic structure of 
most occidental music is still largely poorly understood.  What is
particularly grievous on the other hand is that the choice of music
chosen for experiments mirrors the aesthetic imperialism present in 
the global music industry. 

Furthermore, within Africa, the majority of research on rhythm
curiously involves Ghanian dance drumming. This in itself is peculiar,
as the Ghanian styles are hardly representative of West 
African music. In many respects, any one of the plethora of cultures 
can hardly represent the varied use of rhythms. Again, I suspect that 
the popularity of such studies is strongly linked to the patronization
of Ghanian music schools by the Western academic elite and well-known
composers like Steve Reich, as well as a much stronger push by
Ghanians to export their culture.

\vspace{5mm}
\subsection{Asymmetric Cognitive Clocks}

One useful research effort that appeared in a recent issue of 
{\sl Music Perception} investigated mental models used by expert 
African drummers in the production of polyrhythmic patterns. The 
paper fits into the category of work which pursues computational 
validation of ``culturally initiated beliefs'' about cognitive 
models. (That cultural aspects of a music and its production would 
not even be considered in a computational model is bewildering.)

Experiments were conducted by recording the performance of 
a master Asante drummer's spontaneous patterns and responses to a
computer generated tone. Quantities such as pulse hand allocation
(left or right), pulse stream size (three or four) were varied and 
measured \cite[p. 191]{Magill:97}.  The goal was to determine if the 
African cultural description of the cognitive process involved in 
producing rhythm is valid by computationally modeling the process 
and experimentally verifying its validity. The performance data was also
run through a pulse-ground(PG) model based on the notion of meter as a 
means to check the data against another well understood model \cite[p. 191]{Magill:97}.

The model used in the experiments is an asymmetric time-line-ground(TLG) model 
that represents a computational elaboration of the multilayered
traditional West African understanding that all instruments 
in a percussion ensemble play in relation to a fundamental {\it time-line}.
The asymmetry of the time-line Magill claims doesn't ``reduce to one of
additive meter'' \cite[p. 191]{Magill:97} because much West African
music is founded on the presence of parallel isochronous pulse streams
that regularly cut across the additive subgroups. This is possible
because the cycle lengths of the asymmetric time-line are composite 
numbers (6, 8, 12, 24).  Furthermore, TLG model ``allows for the
presence of unequal clock pulses but assumes that where clock
intervals are of the same duration, the variance of those two clock
pulses will be equal'' \cite[p. 191]{Magill:97}.

In performances by the master Asante drummer, Magill reports that 
the TLG provides a ``convincing'' fit to seven of the eight tests.
The eighth required some modifications to the basic model. 
Magill comments further that the ``asymmetric process'' need not be
based on the single fastest pulse (additive meter) even though the
experimental results can be viewed in this light.

  \ifnum0=\value{mychaptercount}
    \startingpages
    \setcounter{mychaptercount}{1}
  \fi
  \chapter{Description of Algorithms}
\vspace{10mm}

\begin{quote}
  {\it ``Composers shouldn't think too much -- it interferes with their
    plagiarism.''} --- Howard Dietz, American lyricist
\end{quote}

\vspace{7mm}
\section{System Architecture} 
\vspace{5mm}

{\it Riddim} is a system with two high level functions.  
The first takes in audio input and extracts sequences of pertinent 
timing information.  The timing information corresponds to temporal 
gestalt or attack points in the audio. These are points in time 
where there is a large change in the sound pressure level.

The second part of the system takes as input the time sequences 
generated by the first and performs rhythmic timing analyses. 
There are a variety of possible analysis algorithms that try to 
uncover different aspects the multi-dimensional quality that is 
rhythm. The present implementation contains one such analysis -- the
determination per stream of the lowest level pulse. 

The first subsystem works in the following way. A single channel 
audio input is run through a process called Independent Subspace 
Analysis (ISA). For any kind of input that involves two or more instruments or 
sources of sound this analysis technique will successfully separate
each into a different audio channel.  If, for example, we input a
recording of a Latin percussion ensemble, we should get at the output 
a number of different audio channels each containing a different 
instrument present in the recording.

\begin{figure}[thp]
  \begin{center}
    \resizebox{5in}{!}{\includegraphics{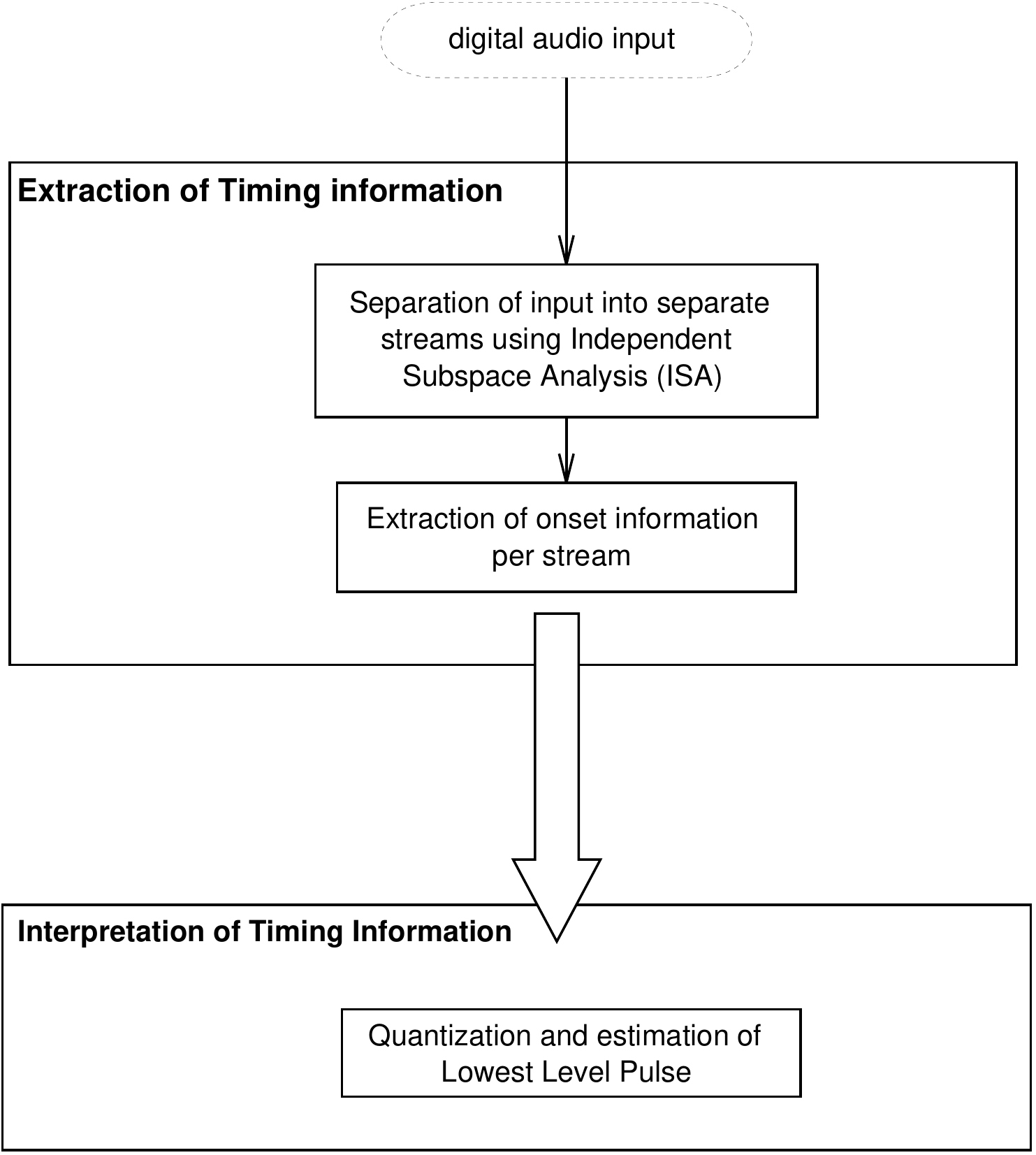}}
    \caption{Riddim: The System Architecture }
    \label{system architecture}
  \end{center}
\end{figure}

Each stream is then passed in succession to an onset detection module.
In this module, a rectify and smooth algorithm is implemented to
prepare the signal for subsequent analyses. Next, a peak detection 
routine finds the points in the stream that correspond to an onset
time or an attack point. The onset detection module will 
return a vector of onset times for each stream that it processes. The 
second high-level subsystem takes in each vector of times from the onset
detection component of the first subsystem, quantizes it and returns the 
lowest level grid for each stream. 

\vspace{5mm}
\subsection{The MATLAB{\texttrademark} Platform}

I chose the MATLAB{\texttrademark} programming language and 
environment as a development platform for my work for several reasons.
First of all it is a common platform for prototyping complex systems,
with widespread support among scientists and engineers, in industry
and academia. It has numerous highly specialized toolboxes for 
mathematical analyses and design in a variety of disciplines. 
This makes it an attractive tool for solving problems quickly, 
more so than a tool to build a commercial application.

Another motivation for using MATLAB{\texttrademark} came from the fact
the implementation of Casey's Independent Subspace Analysis
(ISA) was in MATLAB{\texttrademark}.  Since this was work that formed a 
central part of my application and was one that I intended to use ``as is'' 
with very only minor functional modifications, it made sense to build 
the rest of my application on the same platform.

\vspace{5mm}
\subsection{Native C subroutines via the MEX interface}

However, not all modules were written in MATLAB{\texttrademark}. 
MATLAB{\texttrademark} provides an Application Programmer's 
Interface (API) called MEX that permits external subroutines written in 
C or Fortran to be called from MATLAB{\texttrademark} functions, 
script files or the command line. These special C subroutines are
compiled into MEX-files which are dynamically linked subroutines 
that the MATLAB{\texttrademark} interpreter can load and execute.

Several parts of this work were implemented in C and called via
the MEX interface. These include the bulk of the onset detection 
module and sections of the grid quantization modules.
The main advantages of implementing certain functionality in C is 
speed.  Since MATLAB{\texttrademark} is an interpreted language optimized for 
vector operations, certain constructs like \texttt{for} or \texttt{while} 
loops can be very slow for very large arrays. In such cases, to
increase the overall performance, such operations can be implemented
in C. Another motivation came from the fact that several authors 
sent me demos of their algorithms as MATLAB{\texttrademark}
functions. On one hand it would have been easy to simply use their
code in my implementation. However, to really internalize their work,
I felt that it was important to come up with my own implementation in C.

\vspace{7mm}
\section{Independent Subspace Analysis}
\vspace{3mm}

\subsection{Motivation}

If the goal was to implement a tool that can extract rhythmic 
information from digital audio, what would be the best way to decompose 
the data to facilitate an analysis? How does the brain focus its
attention on a singular element in a sound mixture to be able to pick
out its rhythmic qualities? Does a model of attention really help? 
Is any decomposition or data  reduction even necessary? 

According to Leigh Smith ``The perception  of musically typical rhythms is achieved by 
segregation of the received sound complex into separate streams of 
common sources. It is thereby hypothesized that listeners use timbral, spatial 
localisation, pitch, tempo and other objective differences between 
sound sources to distinguish between independent rhythmic patterns'' \cite[p .85]{Smith:99}.
A first step in implementing a robust perceptual rhythm analysis 
tool is to find a scheme that permits monophonic audio sources to be 
segregated or ``un-mixed'' into their source streams.  In the past, this
problem in the past has been notoriously difficult to solve \cite{Bilmes:93}.
This may be due in part to the characteristic representation of the 
audio signal, the fact that perceptually, sound qualities such 
as timbre and pitch are still evasively difficult to quantify 
accurately and the ``heuristic nature of psycho-acoustic 
grouping rules'' \cite{Casey:2000}.  

First of all, we will discuss some important notions necessary 
to a complete understanding of ISA.

\subsection{The Mechanics of Mixing and Unmixing}

Let us assume that we have two different recordings of an event
involving two instruments. We want to recover the two individual 
instruments which are mixed in varying degrees in the recordings. 

In this simple case, we have two {\it observed} or {\it recorded} 
sounds that we call $y_{1}(t)$ and $y_{2}(t)$, each $n$ samples long. 
We write them in row vector form as, 

\begin{figure}[thp]
  \begin{center}
    \resizebox{4in}{!}{\includegraphics{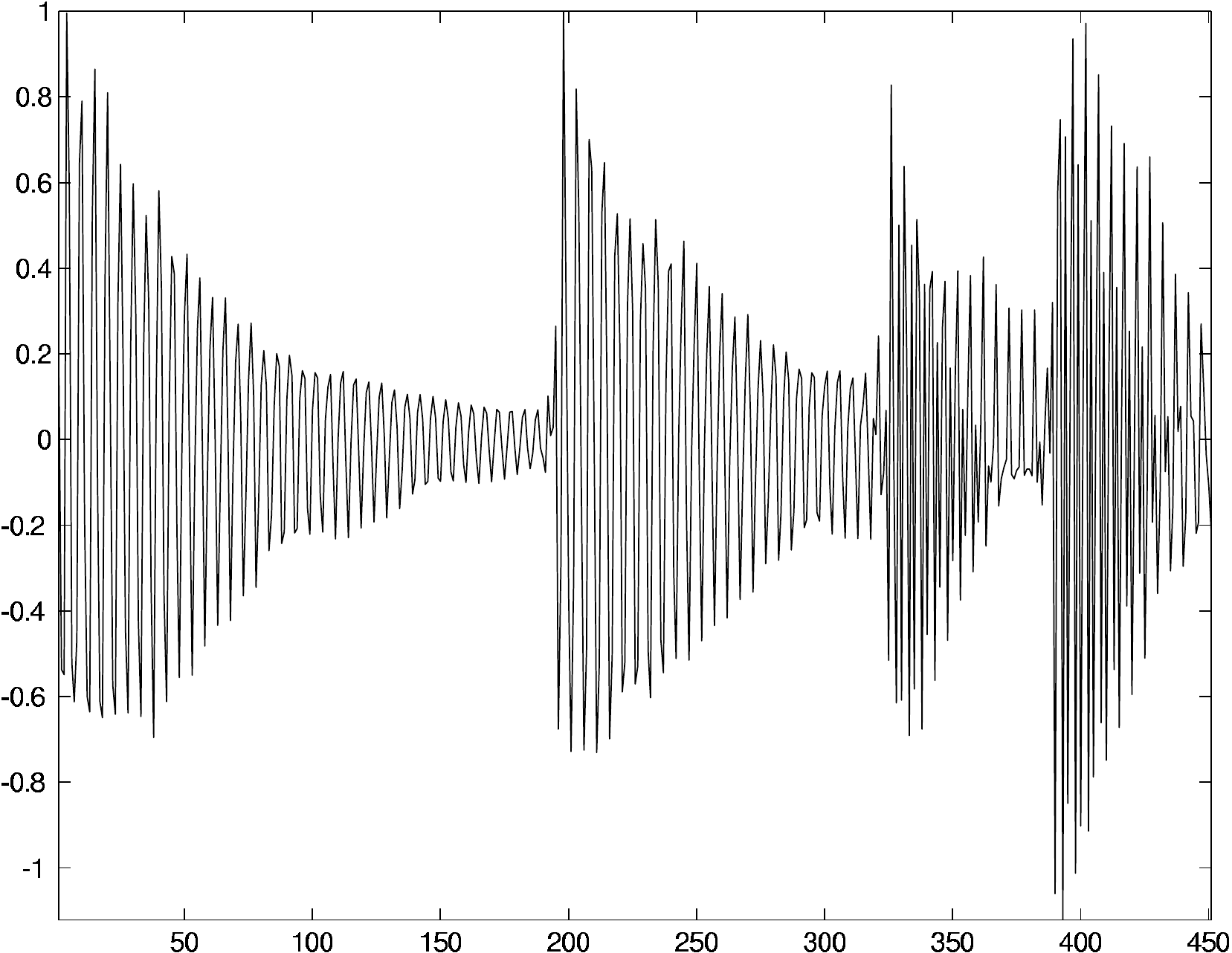}}
    \caption{The waveform (time-series) representation of several
      conga hits 451 samples long.} 
    \label{PDFHistoWaveform}
  \end{center}
\end{figure}

\begin{figure}[thp]
  \begin{center}
    \resizebox{4in}{!}{\includegraphics{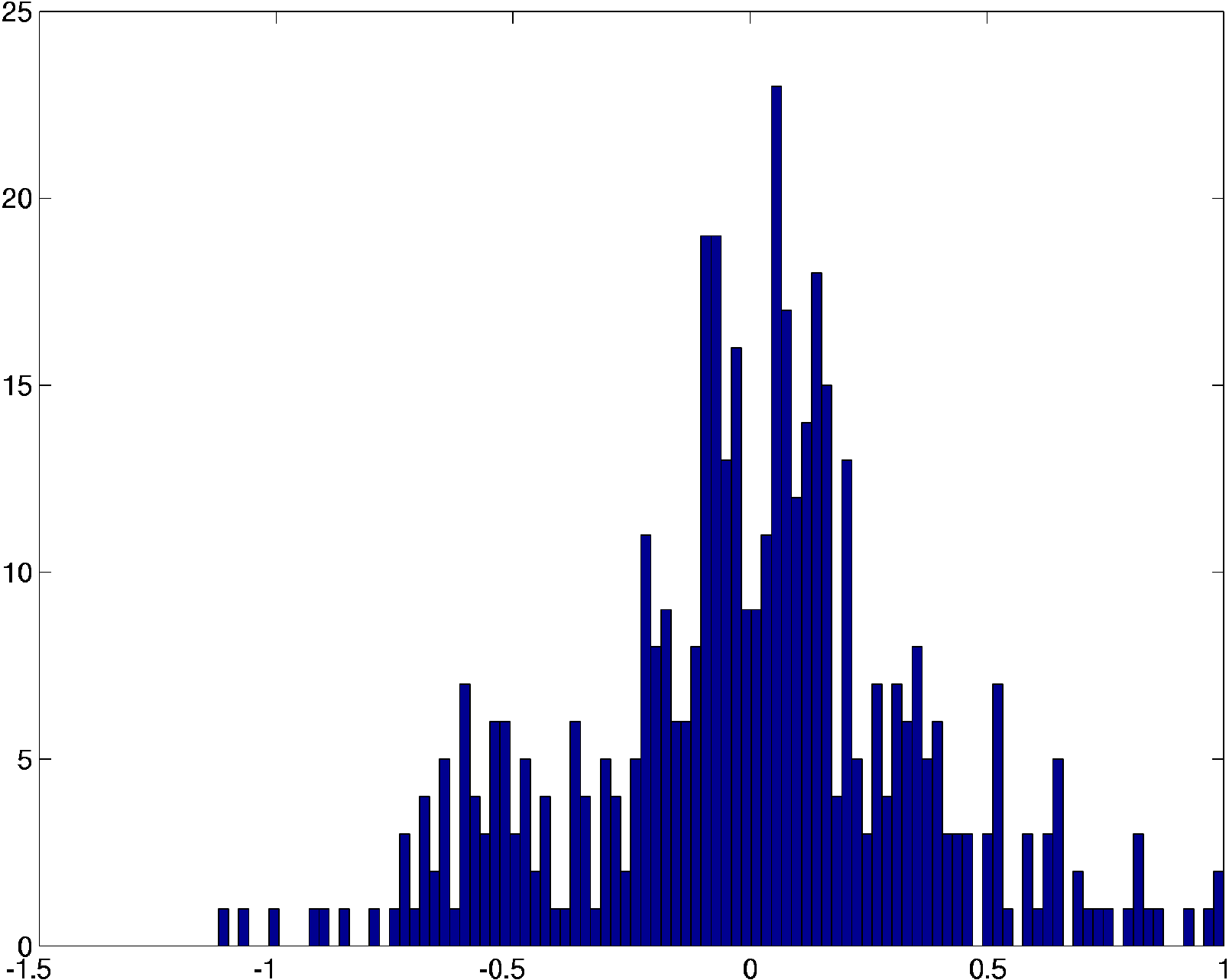}}
    \caption{A histogram representation of Figure
      \ref{PDFHistoWaveform} that approximates a probability density
      function close to a spiky Gaussian.}
    \label{PDFHisto}
  \end{center}
\end{figure}

\begin{equation}
  \label{twoobservedsounds}
  \matrix{
    y_{1}(t) = [y_{1}(1),...,y_{1}(n)] \cr
    y_{2}(t) = [y_{2}(1),...,y_{2}(n)] \cr
    & }
\end{equation}
These observed sounds contain a mixture of sounds from different
instrument sources that we call $x_{1}(t)$ and $x_{2}(t)$ which are also 
$n$ samples long each and are written as,

\begin{equation}
  \label{twolatentsounds}
  \matrix{
    x_{1}(t) = [x_{1}(1),...,x_{1}(n)] \cr
    x_{2}(t) = [x_{2}(1),...,x_{2}(n)] \cr
    &}
\end{equation}
Because of the relatively high sampling rate of the sounds, we can 
represent $x_{1}(t)$ and $x_{2}(t)$ in a histogram of the signals'
amplitudes. The shape of the histograms are a good approximation of 
the probability density functions (PDF) $P(X_{1})$ and $P(X_{2})$ of 
a pair of random variables $X_{1}$ and $X_{2}$. Later we will be
making assumptions about the characteristics of these PDFs.

For notational purposes, we put each pair of vectors into $2
\times n$ matrix so $$ X = \left [
  \matrix{
    x_{1}(t)  \cr
    x_{2}(t)  \cr
    }
\right ] \hbox{ and } Y = 
\left [
  \matrix{
    y_{1}(t)  \cr
    y_{2}(t)  \cr
    }
\right ] $$
[Note that $X$ is not directly related to $X_{1},X_{2}$ defined above.]
There are two assumptions that we take the liberty of making in
discussing how $x_{1}(t)$ and $x_{2}(t)$ are combined to yield $y_{1}(t)$ 
and $y_{2}(t)$. The first assumption is that $x_{1}(t)$ and $x_{2}(t)$
in $X$ are linearly mixed. This means that $Y$ is the product of $X$
and some full rank mixing matrix $M$, 
\begin{equation}
  \label{mixing}
  Y = MX
\end{equation}
where the $M$ realizes the linear mixing.  Thus to recover $X$ given
$Y$ reduces to left multiplying $Y$ by the $M^{-1}$,
\begin{equation}
  \label{unmixing}
  X = M^{-1}Y
\end{equation}
In our situation, pulling recordings from CDs or vinyl, we have 
the mixture $Y$ and we don't know $M$. Thus to recover the sources $X$
we must try to estimate $M^{-1}$ from the mixture $Y$. 
To help this estimation, a second assumption is made. This is that 
random variables $X_{1}$ and $X_{2}$ (introduced above), whose PDFs are
modeled by the histograms of $x_{1}(t)$ and $x_{2}(t)$ respectively, are 
{\it independent}.  This means that their joint probability
distribution given by,
\begin{equation}
  \label{independence}
  P(X_{1},X_{2}) = P(X_{1}) \cdot P(X_{2})
\end{equation}
is separable. Plotting the histogram for two sound sources that are
independent with uniform distributions we see that ideally their 
joint distribution is square-like. On the other hand,
plotting the histograms for two highly correlated sounds, we see that
their joint distribution tends to run along an axis. 

\begin{figure}[thp]
  \begin{center}
    \resizebox{5in}{!}{\includegraphics{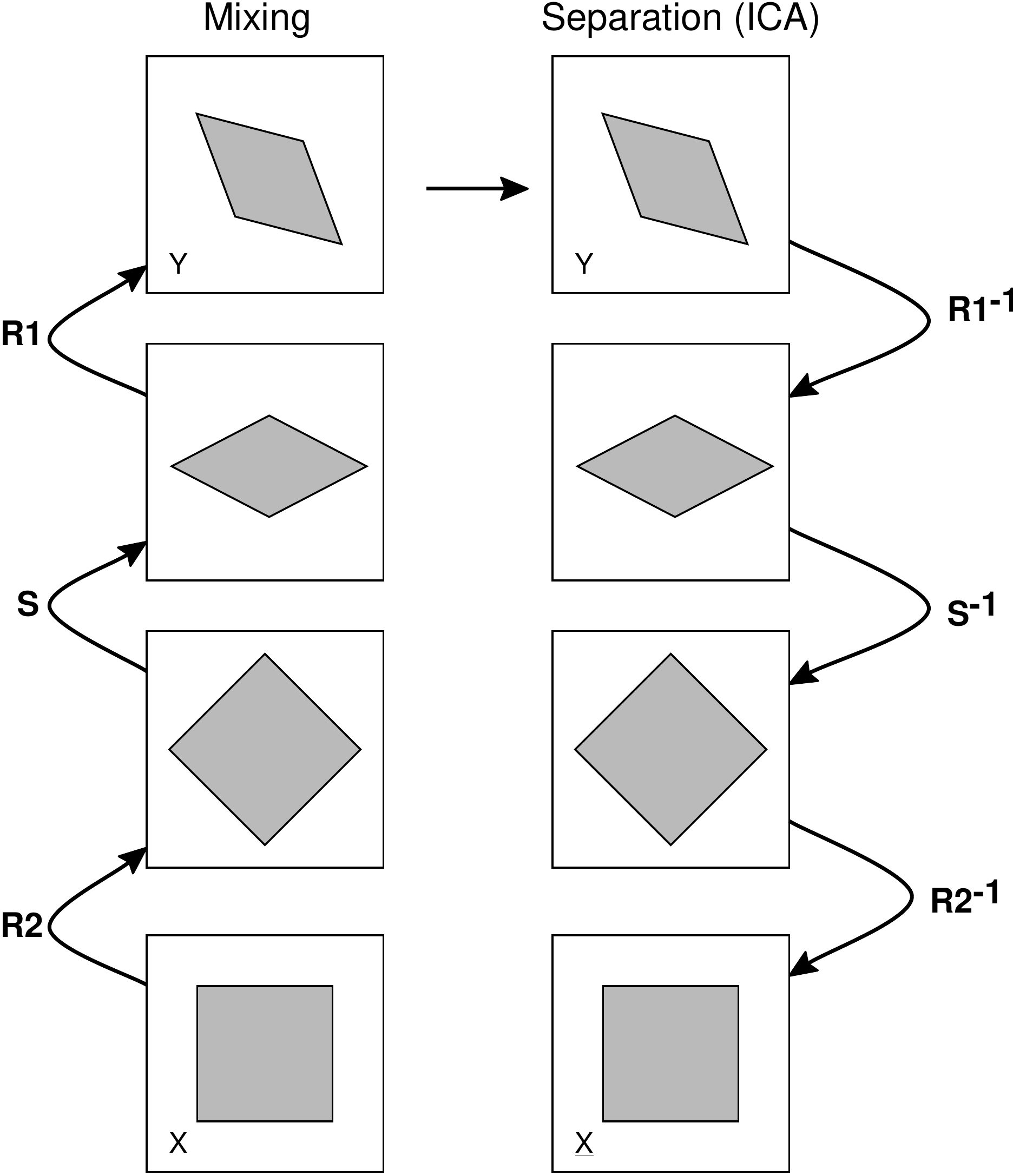}}
    \caption{From \cite{Farid:99}, the bottom left shows an
      ideal joint probability density function of two independent
      signals in $X$. The linear mixing of the signals in equation
      \ref{mixing} transforms the distribution by rotating 
      it with $R_{2}$, scaling it with $S$ and rotating it with
      $R_{1}$. ICA tries to estimate $M^{-1}$ by finding rotation
      matrices $R_{1}^{-1}$ and $R_{2}^{-1}$, and scaling matrix
      $S^{-1}$ that transforms the joint PDF of the mixed signals back to a
      square.}
    \label{MixingUnmixing}
  \end{center}
\end{figure}

\vspace{5mm}
\subsubsection{Decomposing the Mixing Matrix}

From equation \ref{mixing}, one can gain some intuition into the
mechanics of the mixing procedure by diagonalizing the matrix $M$.
Using the Singular Value Decomposition (SVD), $M$ can be expressed as 
\begin{equation}
  \label{SVD}
  M = R_{1}SR_{2}
\end{equation} where $R_{1}$ and $R_{2}$ are orthonormal matrices and
$S$ a diagonal matrix \cite{Farid:99}. Figure \ref{MixingUnmixing}
shows that $M$ applied to the idealized joint distribution of a pair of independent
signals can be seen as matrix $R_{2}$ performing a rotation, diagonal
matrix $S$, a scaling, and $R_{2}$ a final rotation to yield the 
joint PDF of the mixed signals.

Thus estimating $M^{-1}$ from equation \ref{unmixing} reduces to
finding rotation and scaling matrices that undo the mixing operations.

\vspace{5mm}
\subsection{Principle Component Analysis}

\begin{figure}[thp]
  \begin{center}
    \resizebox{3.5in}{!}{\includegraphics{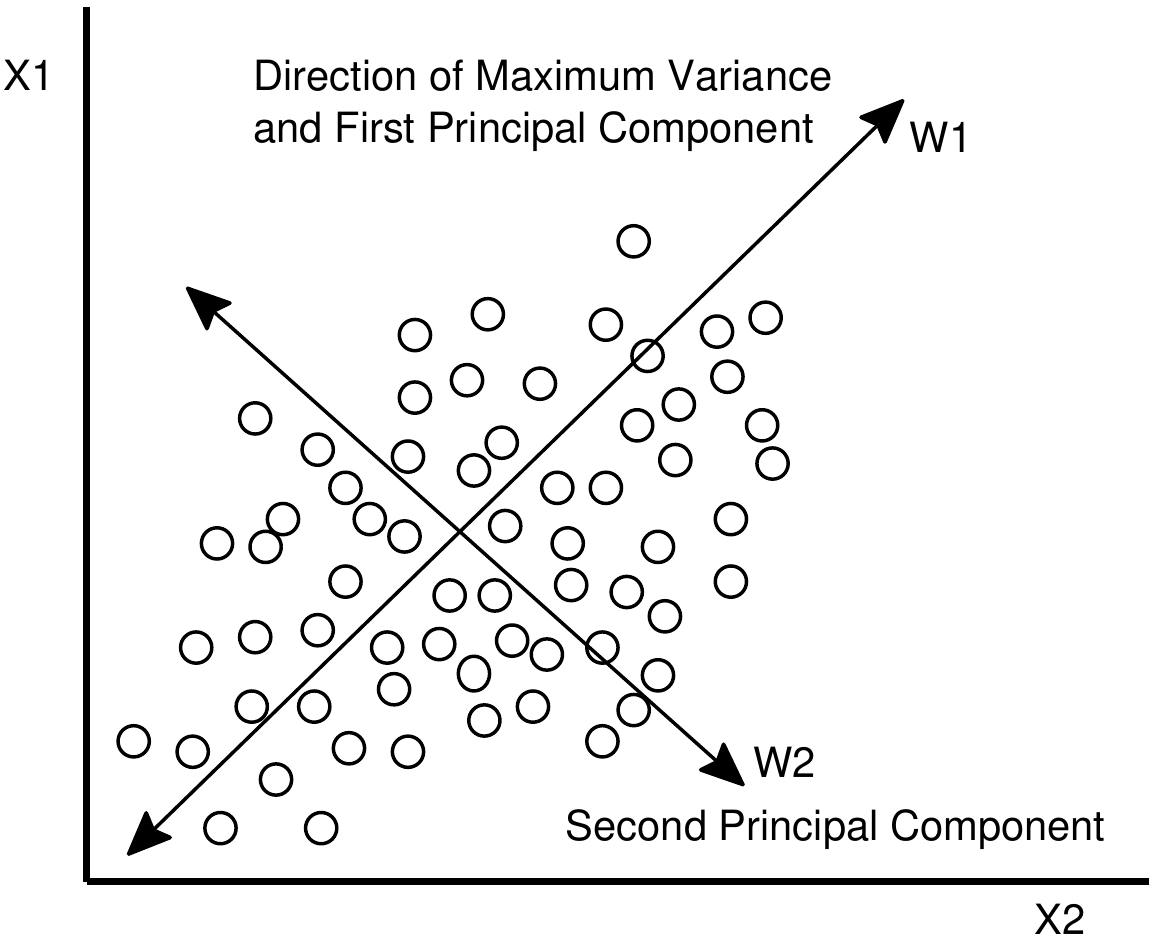}}
    \caption{Principal Component Analysis: $Y1$ and $Y2$ show
      directions of the first two principle components \cite{Cook:97}.}
    \label{PCA}
  \end{center}
\end{figure}

The matrices, $R_{1}^{-1}$ and $S^{-1}$, that perform the
first two inverse operations respectively from Figure
\ref{MixingUnmixing} are obtained via a technique called Principal 
Components Analysis (PCA). PCA inspects the variance structure of 
the data looking for components that account for most of the variation
in the data. It is about the axis of greatest variance that the joint
PDF of the mixed signals (in the upper right corner of Figure \ref{MixingUnmixing}) is
first rotated. This matrix $R_{1}^{-1}$ is obtained first by
calculating the covariance matrix of $X$, written as,
\begin{equation}
  \label{covariance}
  {\bf C}  = XX^{T}
\end{equation}
Next, ${\bf C}$ is diagonalized using SVD to yield three new matrices \cite{Strang:93},
\begin{equation}
{\bf C} = UDV^{T}
\label{SVDCovariance}
\end{equation}
In a zero mean data set, the axis of maximum variance is given by the 
eigenvector corresponding to the first eigenvalue of the covariance matrix ${\bf
C}$. Refer to Figure \ref{PCA}. Thus from equation \ref{SVDCovariance}, $R_{1}^{-1} = V^{T}$. 

The next transformation matrix we must estimate as part of the
separation process from Figure \ref{MixingUnmixing} is the diagonal
scaling matrix $S^{-1}$. In practice it is given by the diagonal
matrix $D$ in equation \ref{SVDCovariance}, so $S^{-1} = D$. 

The final rotation matrix $R_{2}^{-1}$ from Figure
\ref{MixingUnmixing} is a bit more difficult to obtain. 
Typically, the angle that transforms the diamond joint PDF into a
square occurs where the kurtosis or fourth-order cumulant of the joint
PDF is minimized. So an error function of the orientation angle of the
joint PDF is defined and local minima of this function across all
orientations angles give a number of possible candidates for
the final rotation matrix $R_{2}^{-1}$ \cite{Farid:99}. 

\vspace{3mm}
\subsubsection{Statistical Independence and Mutual Information}

To narrow down the candidates matrices to one, we
must go back to an initial assumption from equation
\ref{independence}. Here we claimed that the joint PDF of two 
random variables is the same as the product of their PDFs
when the variables are independent. Generalizing equation
\ref{independence} for $N$ variables instead of two, statistical independence
is given by,
\begin{equation}
\label{statisticalIndependence}
P(X_{1},\cdots,X_{N}) = \prod_{i = 1}^N P(X_{i})
\end{equation}
Statistical independence is achieved when the distance between the
joint PDF and the product of the marginal PDFs is minimized.
To calculate distances between PDFs, a measure like the
Kullback-Leibler divergence is used, given for two PDF $P(X), Q(X)$
by, 
\begin{equation}
\label{generalKL}
K(P||Q) = \int P(X)log \Biggl( { P(X)\over Q(X)} \Biggr) dX
\end{equation}
Furthermore, if the probability density function of a mixture factors
into the product of the marginal densities (statistical independence)
the mutual information between the output joint density and the
component densities is said to be zero. 

Thus, by applying each of the candidate rotation matrices obtained
from the local minima of the error function minimizing the kurtosis 
of the joint distribution, the mutual information is calculated. 
The matrix that has the lowest measure of mutual information between 
the joint PDF and the product of the marginal PDFs obtained from its
application to the joint PDF becomes final rotation matrix
$R_{2}^{-1}$ in Figure \ref{MixingUnmixing}.

In practice, though as was mentioned above, the PDFs are approximated 
by the histogram representation of the mixed and individual signals. 

\vspace{5mm}
\subsubsection{Summary}

\begin{figure}[thp]
  \begin{center}
    \resizebox{3.75in}{!}{\includegraphics{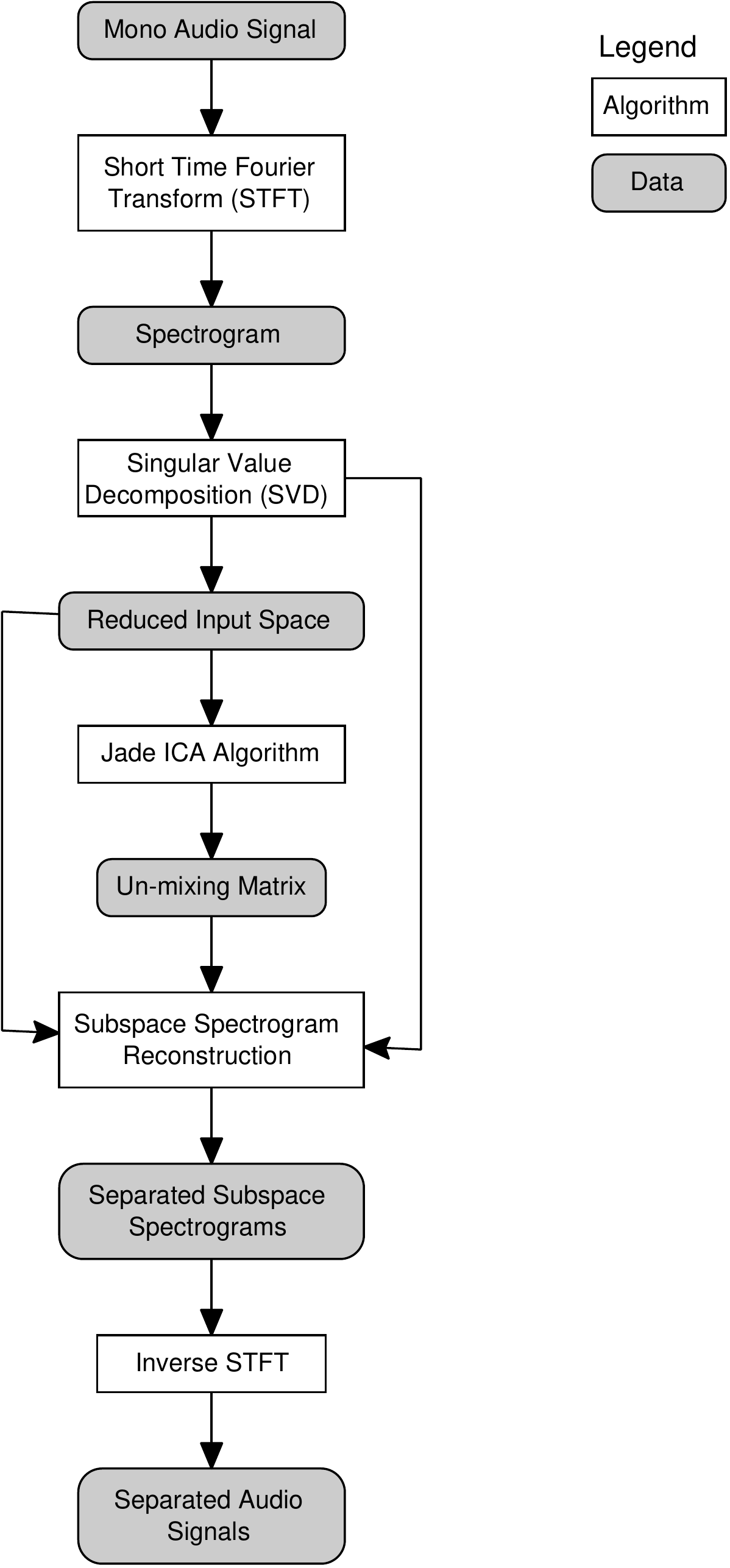}}
    \caption{High Level Functional Blocks of the ISA Algorithm}
    \label{ISAGeneral}
  \end{center}
\end{figure}

The notion of independence is of central importance to this separation
scheme. Ideas on independence are heavily influenced by Gestalt 
grouping rules and Bregman's views on stream segregation in auditory scene
analysis \cite{Bregman:90}\cite{Casey:2000}. Described above are the
mechanics of one approach to solving the canonical separation problem
using ICA. There are a variety of algorithms in the literature that
use ``higher-order statistics, minimum mutual information or maximum
entropy in their solutions'' \cite{Knuth:98}.

For a more in-depth discussion on independence, mutual information,
ICA algorithms and their derivations, please refer to \cite{Knuth:98}\cite{hyvarinen:99survey}\cite{smaragdis:97}\cite{Farid:99}\cite{Casey:2000}.

\vspace{5mm}
\subsection{Independent Subspace Analysis}

Casey's innovation in ISA was the idea to take a mono signal (that
ordinarily cannot be un-mixed directly using ICA) and perform
a change of basis operation before employing canonical ICA techniques.
A mono signal of size $1 \times N$ is first projected onto a new bases of sines and
cosines using a windowed Short Time Fourier Transform (STFT) to yield
a spectrogram of size $n \times m$. This new multidimensional manifold
is composed of $m$ time slices each containing $n$ frequency bins.
\begin{figure}[thp]
  \begin{center}
    \resizebox{5in}{!}{\includegraphics{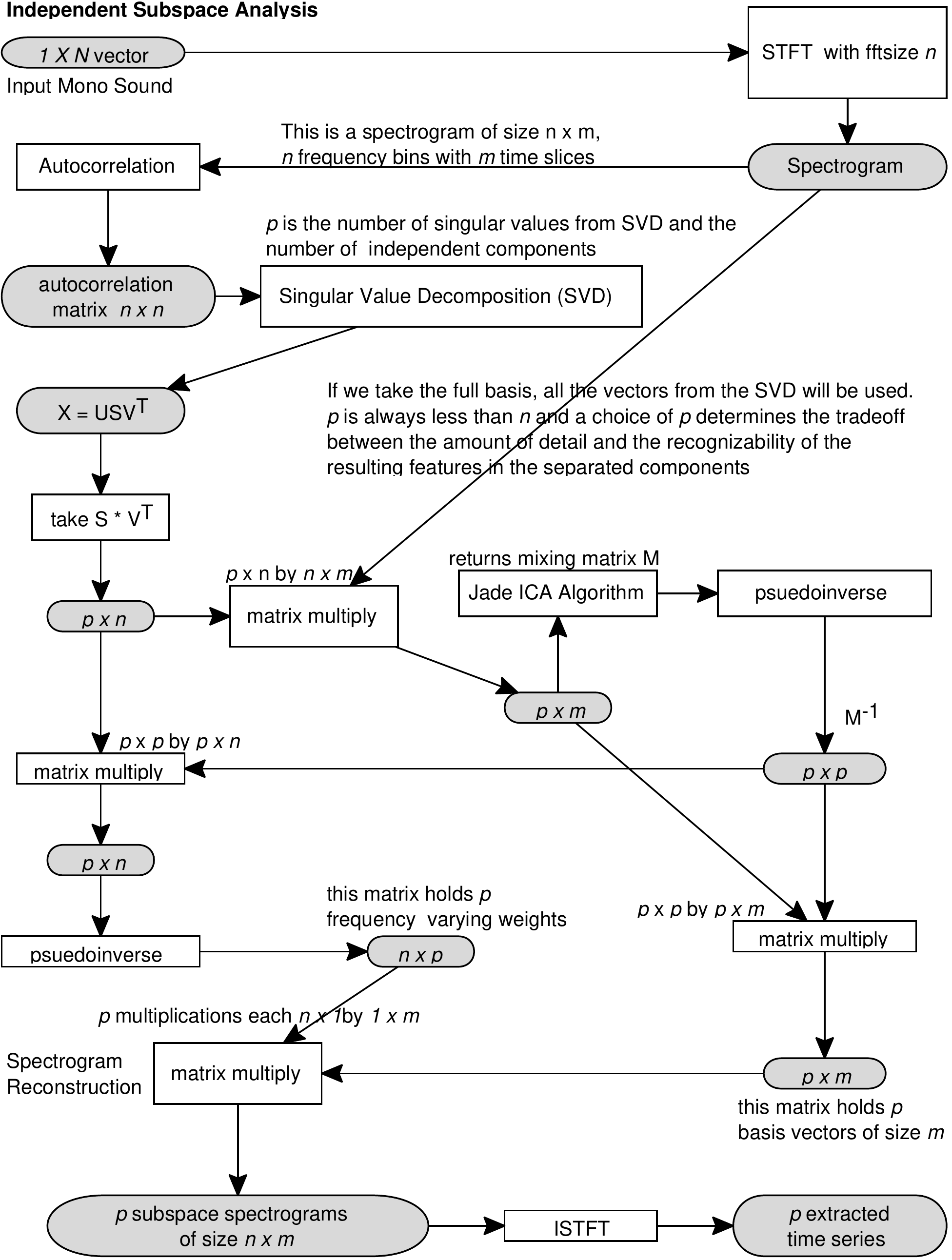}}
    \caption{Details of the ISA algorithm}
    \label{isaflow}
  \end{center}
\end{figure}
The general approach is to do a dimension reduction on this
high dimension space to obtain a reduced set of vectors spanning the
input space. This is accomplished by performing SVD on the covariance
matrix of the input spectrogram. The input spectrogram is then
projected onto this new basis to yield a dimension-reduced input space.
The input basis vectors are then fed to the Jade\footnote{Jade is an
ICA algorithm by Jean-Fran{\c c}ois Cardoso} algorithm which returns
the mixing matrix $A \approx M^{-1}$.  The un-mixing matrix is then 
multiplied against the dimension-reduced basis vectors from the 
spectrogram projection to yield the independent components oriented in time.  

\begin{landscape}
  \begin{figure}
    \centering
    \includegraphics[width=7in]{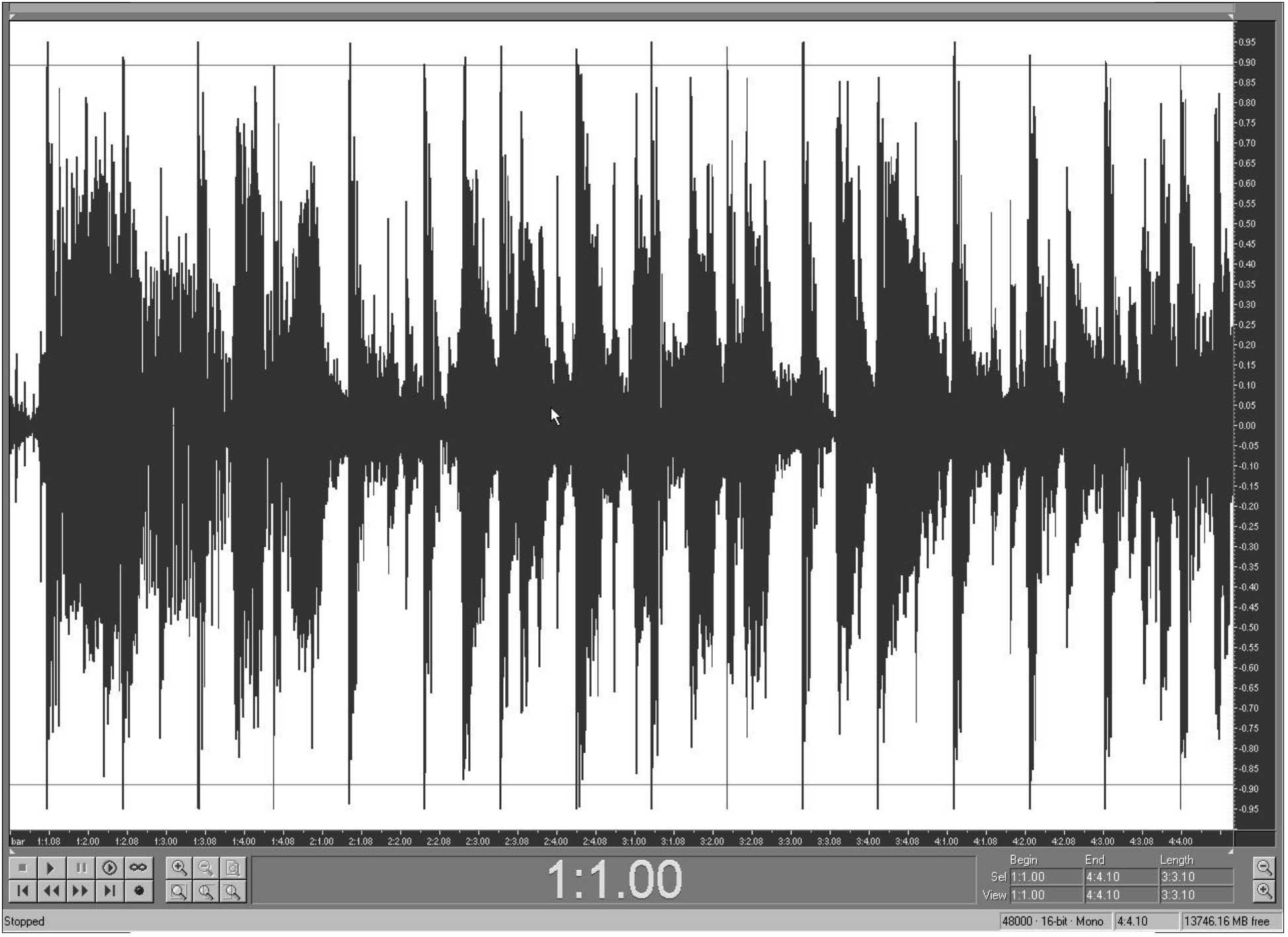}
    \caption{A waveform of an eight second excerpt of a recording of
      Soukous music from the Congo}
    \label{zingzongoriginal}
  \end{figure}
\end{landscape}

\begin{landscape}
  \begin{figure}
    \centering
    \includegraphics[width=7in]{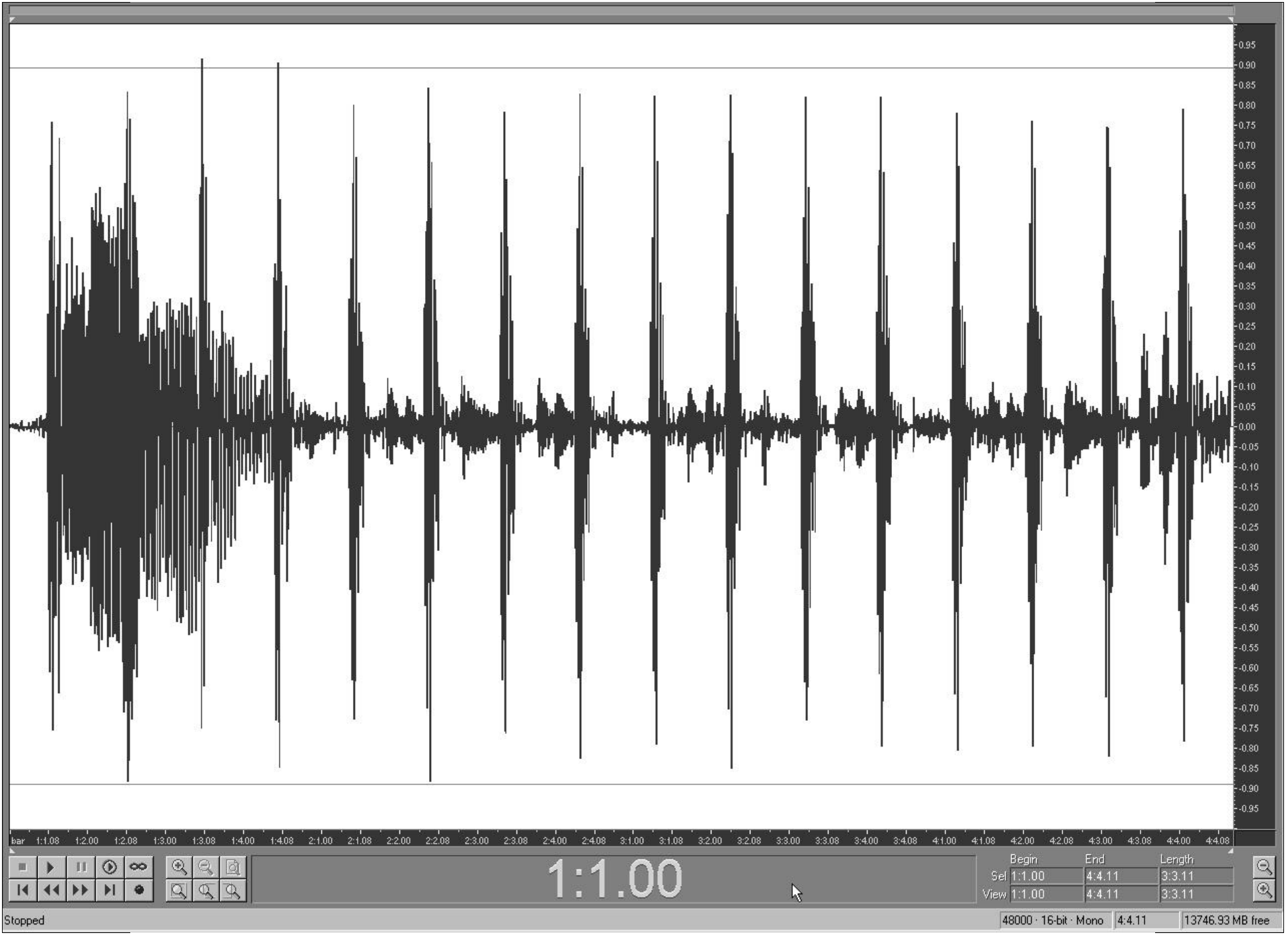}
    \caption{The extracted kick drum from the Soukous music}
    \label{zingzongkick}
  \end{figure}
\end{landscape}

\begin{landscape}
  \begin{figure}
    \centering
    \includegraphics[width=7in]{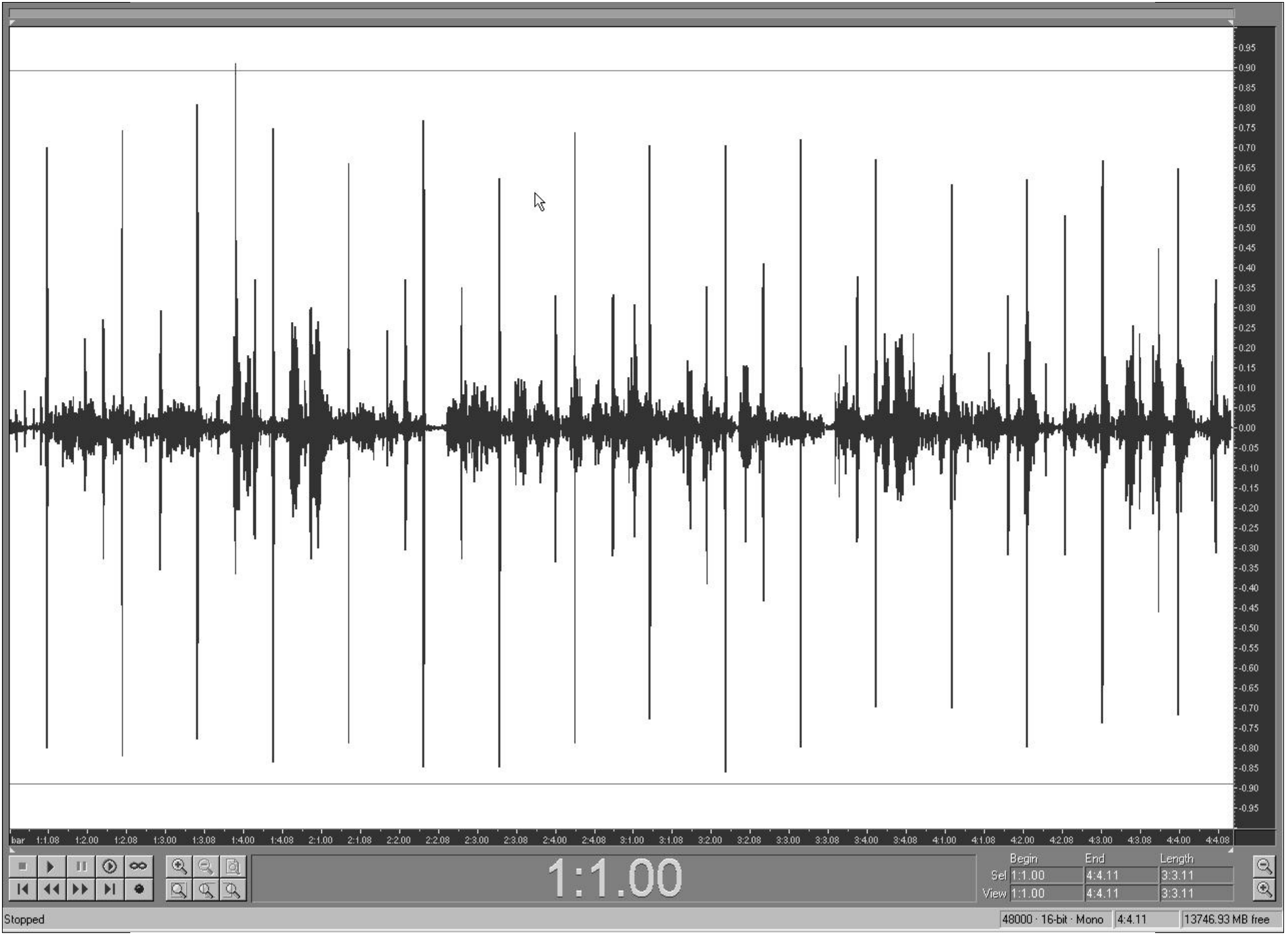}
    \caption{The extracted snare drum from the Soukous music}
    \label{zingzongsnare}
  \end{figure}
\end{landscape}

\begin{landscape}
  \begin{figure}
    \centering
    \includegraphics[width=7in]{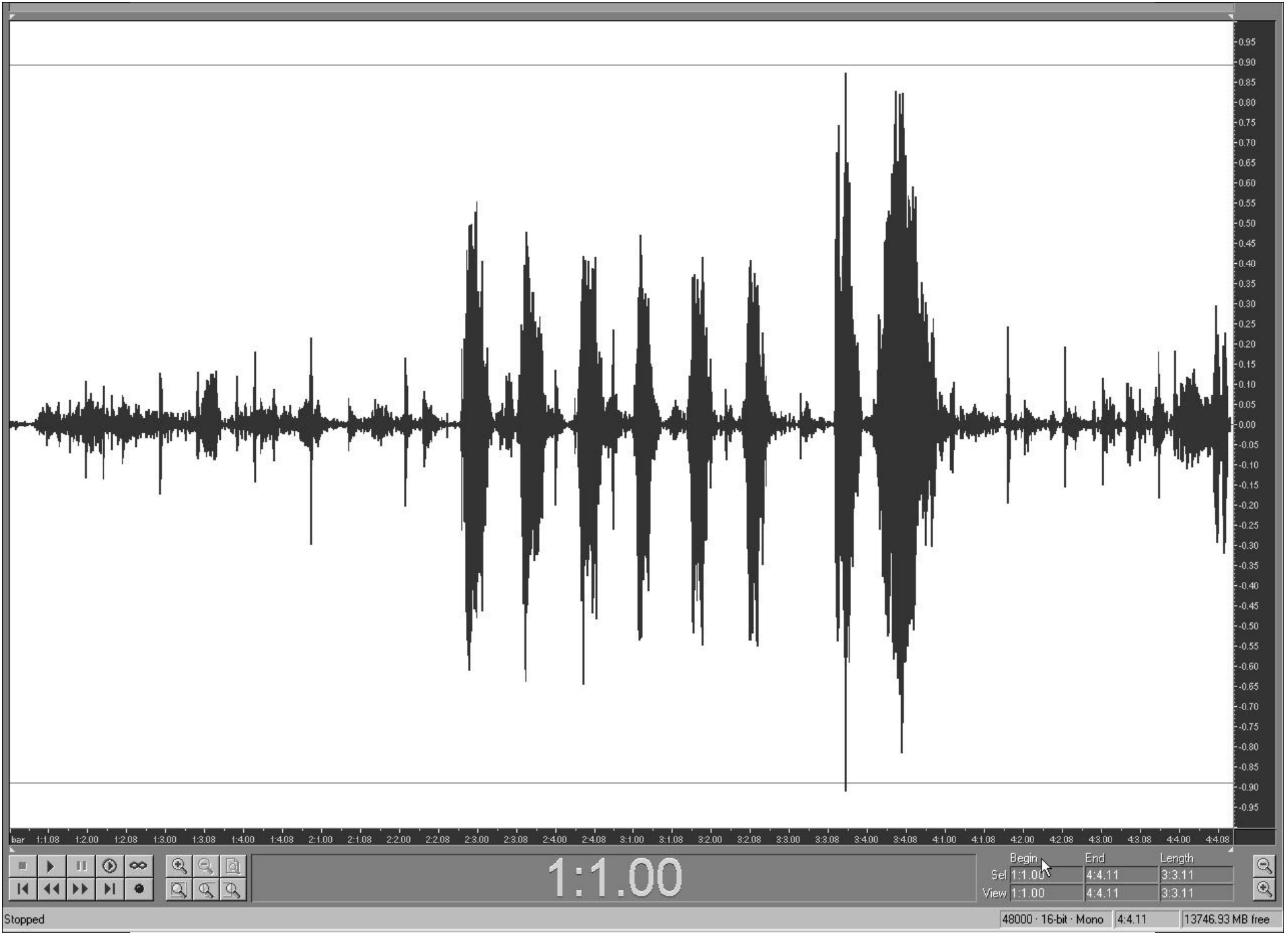}
    \caption{The extracted human vocal chants from the Soukous music}
    \label{zingzongvox}
  \end{figure}
\end{landscape}

\begin{landscape}
  \begin{figure}
    \centering
    \includegraphics[width=7in]{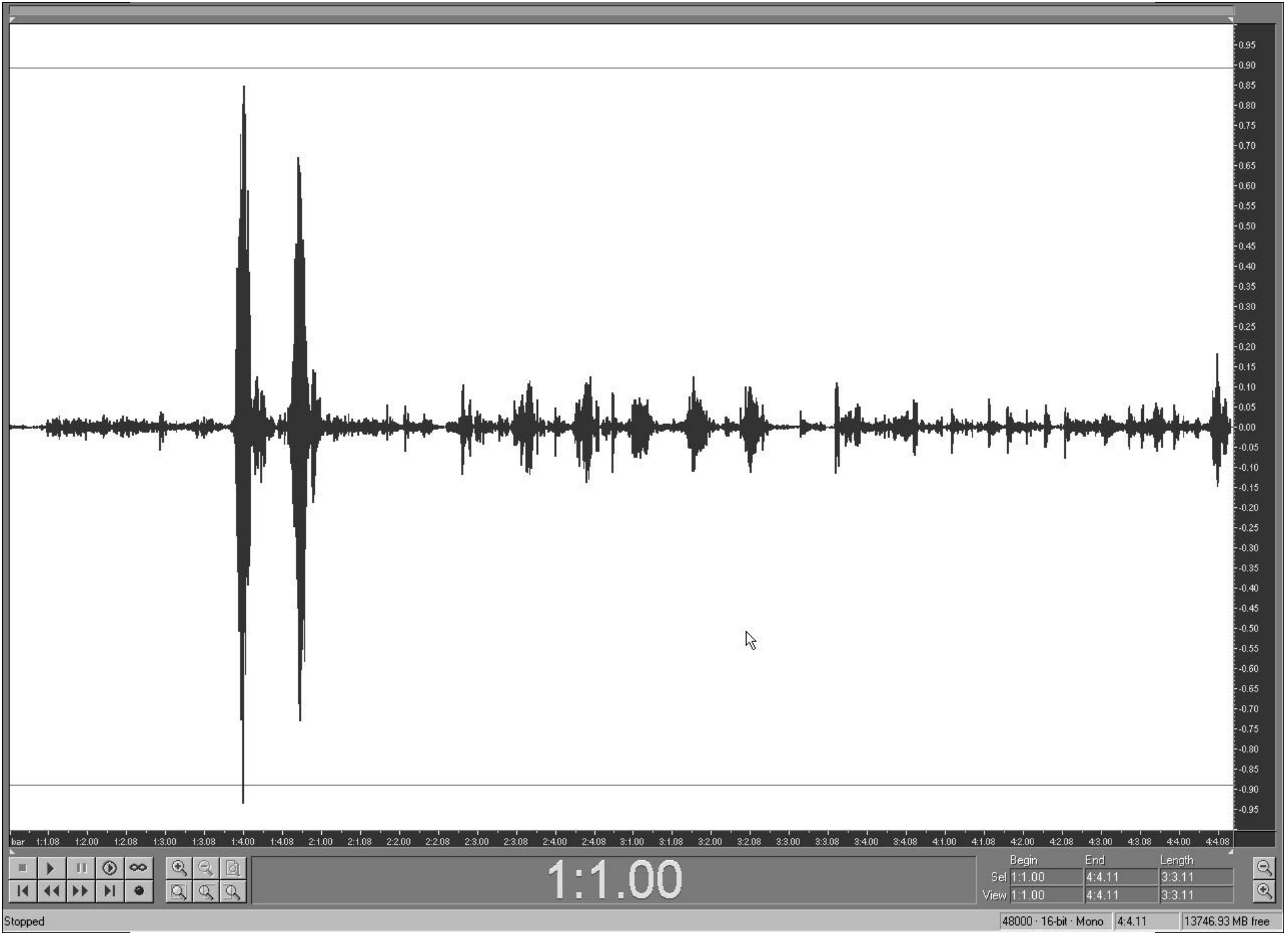}
    \caption{The extracted whistle blow from the Soukous music}
    \label{zingzongwhistle}
  \end{figure}
\end{landscape}

Next, frequency varying weights for each independent component are 
calculated by multiplying the un-mixing matrix against the SVD reduced
vectors. Together, the basis vectors and the frequency varying weights
are combined to yield individual subspace spectrograms corresponding
to independent marginal distributions within the original
time-frequency space. These spectrograms are then passed through a 
windowed inverse STFT (ISTFT) to yield the time domain audio signals 
corresponding to the extracted streams. 

\vspace{7mm}
\section{Detecting Onset Points}
\vspace{3mm}

The next subsystem in the processing chain extracts onset timing
information from each of the audio streams passed to it. The onset detection
algorithm used was taken from a paper on sound segmentation by 
Anssi Klapuri \cite{Klapuri:99} at Tampere University of Technology 
(TUT) in Finland.

Onset detection is based on the fact that humans are built to detect real-world
structure by detecting changes along physical dimensions, representing
the changes as relations \cite{Jones:76}.

Researchers in sound segmentation and onset detection agree that
a robust system should imitate the human auditory system by treating
frequency bands separately and combining onset information from each
band at the end of the analysis. 
\begin{figure}[thp]
  \begin{center}
    \resizebox{3in}{!}{\includegraphics{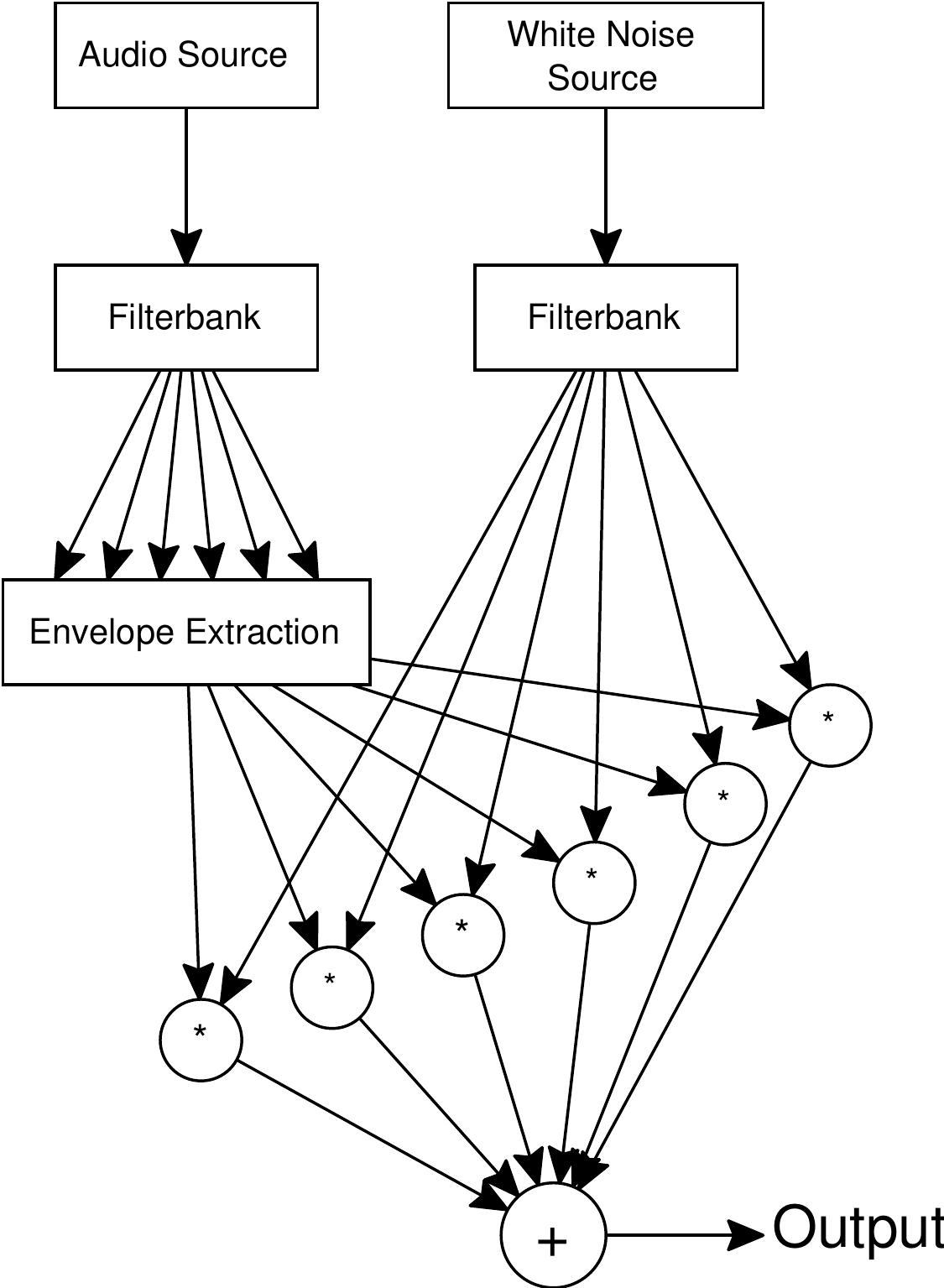}}
    \caption{``Amplitude Modulated'' white noise from a musical signal
      has the same rhythmic percept as the signal}
    \label{AMod}
  \end{center}
\end{figure}
The first detection systems tried to analyse the amplitude envelope of
the audio signal as a whole without any band filtering.  Since the
results were not very accurate, researchers moved to concurrent
analyses in different frequency bands combining the various
band results at the end \cite{Klapuri:99}\cite{Scheirer:98tempo}.
In a sense, separately processing frequency bands crudely imitates the human
auditory system's tuning curves. This approximation becomes
unnecessary when individual streams are separated. Thus, in my
implementation of Klapuri's onset tracking algorithm, I do not preprocess
the extracted streams through any filterbanks, but analyse each stream as is.  

Another psychoacoustic principle that is employed to simplify 
the onset detection algorithm deals with the perceptual rhythmic 
proximity of an amplitude modulated noise signal to the original signal. 
One can construct an amplitude-modulated noise signal by passing 
a white noise signal and a musical signal through the same set of 
filter banks. The amplitude envelope of the output of each of the 
music signal filter banks is used to control the corresponding noise 
band amplitude. The resulting noise signals are then summed together
to form an output signal. Refer to Figure \ref{AMod}.

Scheirer reports that ``when an audio signal is divided into  
at least four frequency bands and the corresponding bands of a 
noise signal are controlled by the amplitude envelopes of the 
musical signal, the noise signal will have a rhythmic percept 
which is significantly the same as that of the original signal'' \cite{Scheirer:98tempo}. 
The importance of this is that since ``the only thing preserved in this
transformation is the amplitude envelopes of the filter bank outputs,
it stands to reason that only this much information is necessary to
extract pulse and meter from a musical signal; that is, algorithms for
pulse extraction can be created which operate only on this much input
data and ``notes'' are not a necessary component for hearing rhythm'' \cite{Scheirer:98tempo}. 

The importance of this psychoacoustic simplification is that the
task of finding rhythmically significant time points in a section of
audio reduces to finding the same points on a drastically smaller data
set, i.e. the smoothed amplitude envelope of that same section of
audio. This smoothed version of the original is attained via a
rectification and smoothing algorithm.

\begin{landscape}
  \begin{figure}
    \centering
    \includegraphics[width=7in]{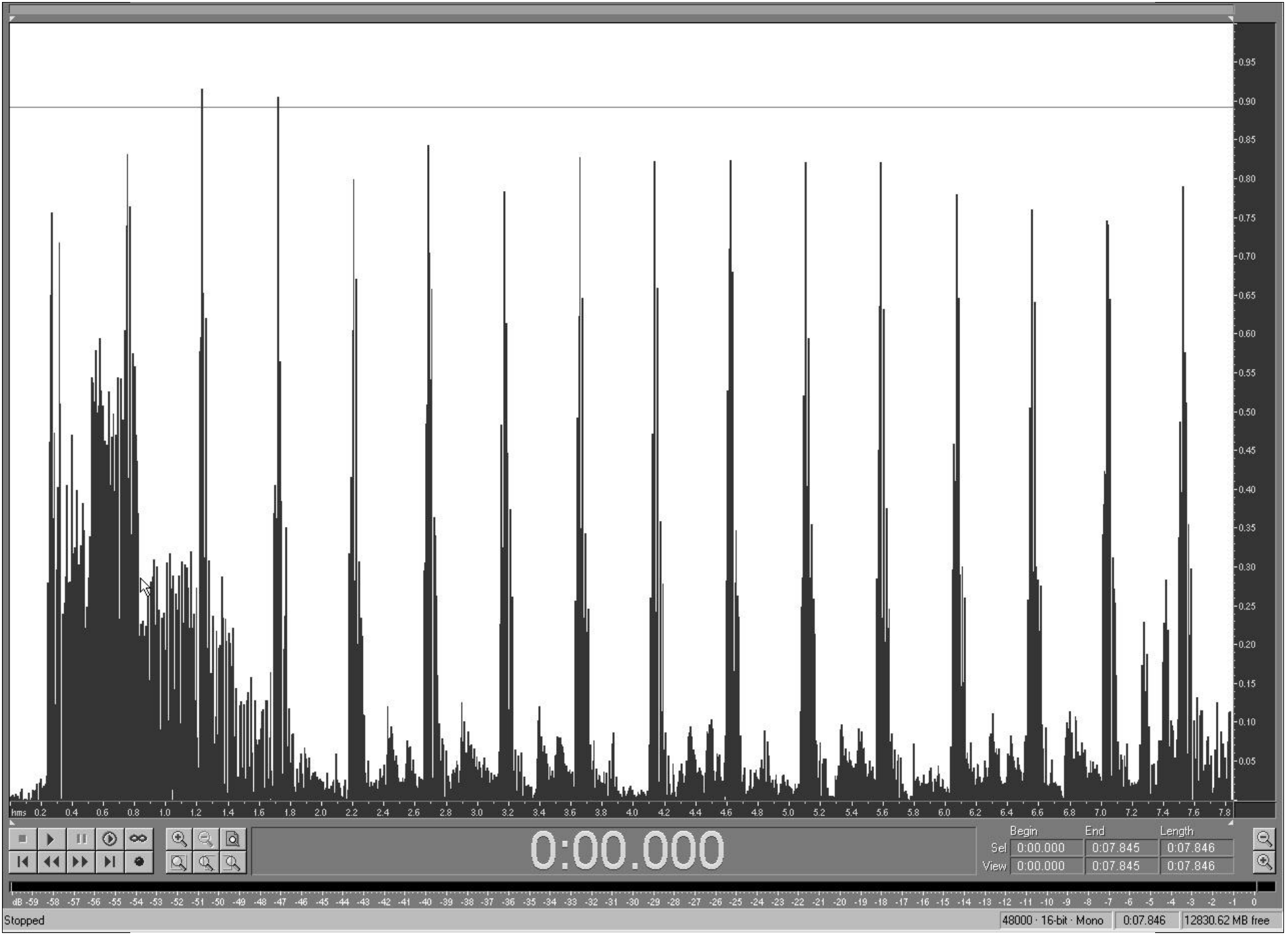}
    \caption{The soukous kick drum after its been half-wave rectified}
    \label{kickrect}
  \end{figure}
\end{landscape}

\vspace{5mm}
\subsection{Rectification and Smoothing}

From an audio signal with potentially cumbersome imperfections our task is to prepare 
the signal for a robust peak searching algorithm by first applying a
rectify-and-smooth algorithm proposed by Scheirer \cite{Scheirer:98tempo}.
First, the input audio stream is half-wave rectified. This means that all the
negative bits of the signal cycle are converted to zero. Refer to
Figure \ref{zingzongkick} and \ref{kickrect} for a graphical representation of
what occurs before and after rectification.

At this stage, the rectified signal is decimated. This involves a two
step process in which the signal is low pass filtered with a sixth
order Chebyshev Type I lowpass filter with a cutoff frequency,
$$w = {.8(\hbox{ samplerate}) \over 2R}$$ where $R$ is the decimation factor. The
second part of the decimation process is the actual resampling of the
signal at $1/R$ its original rate.  Decimation is implemented to allow
all further operations on the data to be more computationally
efficient, since there is minimal data loss for signals with sample
rates as high as $44100 \ samples/second$.

The decimated signal is then convolved with a 200 msec raised cosine 
window.  The window's discontinuity in time (i.e. the start and end 
points are not at the same level) means that its Fourier transform 
has a non-linear phase response. This means that lower frequencies are 
passed through with a larger delay than higher ones.  The window
has low pass filter qualities with a -10dB response at 10Hz and $6
dB/octave$ rolloff afterwards. It performs a kind of energy integration
similar to what occurs in the ear by weighting the most recent inputs
and masking rapid amplitude modulations in the signal.  The net effect
of convolving the input data with this window is that all the rough 
edges, discontinuities, rapid modulations and other ``imperfections'' 
in the signal are smoothed and we are left with just the amplitude 
envelope of the original signal \cite{Scheirer:98tempo}.

\vspace{5mm}
\subsection{Amplitude Envelope Peak Detection}

Once the signal has been suitably smoothed, it is now time to search
for peaks or sudden changes in signal intensity or representative
sound pressure level that would suggest the presence of an attack
point or onset.  Most algorithms from the literature accomplish this 
task in one way or another by examining peaks in the first order 
difference (FOD) function of the original signal, which correspond 
to the points on the signal with the steepest positive slope. For a 
time domain signal S(t) the first order difference function FOD can 
be defined as: 
\begin{equation}
FOD(t) = {1 \over \Delta t} (S(t) - S(t-1))
\label{FOD}
\end{equation}

\begin{figure}[thp]
  \begin{center}
    \resizebox{4in}{!}{\includegraphics{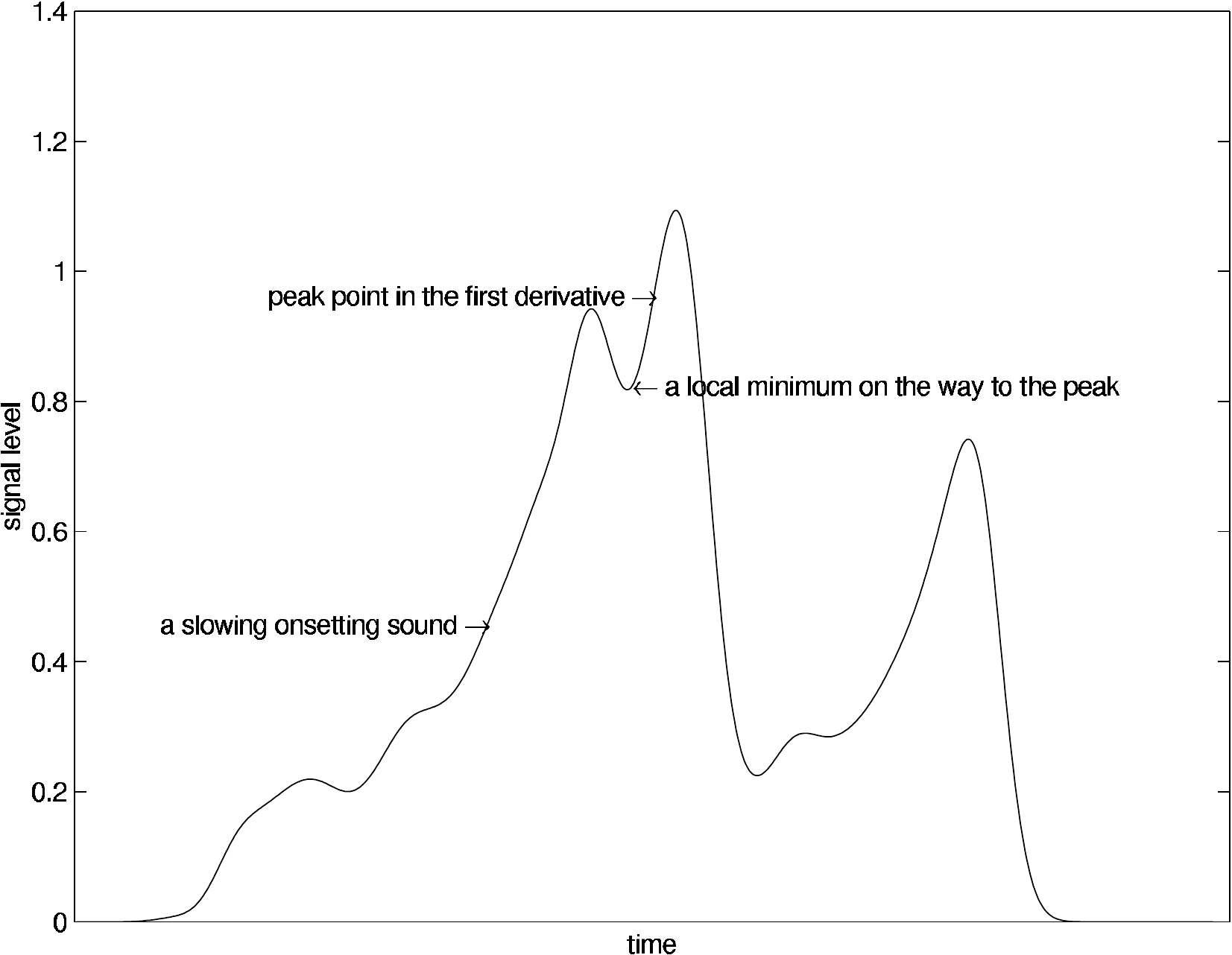}}
    \label{A Slowly Onsetting Sound}
    \caption{When using peaks in the {\sl first order difference
        function} to determine onset times, the estimation is
      sometimes late with slowly onsetting sounds and local maxima may
      be picked as separate onsets.}
  \end{center}
\end{figure}

Klapuri's algorithm is different in several ways that directly address
two problems when using the peaks of the FOD function. First of all,
researchers at TUT noticed that the first derivative measures the
signal loudness well, but the maximum value of the signal's slope does not
correspond directly to the start time of the onset. This is due to the
fact that if the sound takes some time to rise to its maximum value, 
the point at which there is a maximum slope might be late in relation 
to the physical onset of the sound. Note, I refer here to the {\sl
physical onset} not the perceptual onset. The distinction is important
since the physical onset is the one that would be used in a tool like
Recycle{\texttrademark} for splicing and rearranging loops. The
perceptual onset, which is where the human would subjectively mark as the
beginning of the event, may not necessarily coincide with the physical
onset. If the sound rises slowly, its perceptual onset may be
significantly later than the physical onset. Additionally, many sounds
as they rise from do not monotonically increase. There are usually
many local maxima and minima throughout the attack point that may 
be confused as separate attacks \cite{Klapuri:99}.

To handle these problems, Klapuri suggests using the {\sl relative
difference function} (RDF). The RDF in layman's terms is the ratio of the FOD 
to the original signal. It measures the amount of change in
a signal relative to its level and is equivalent to the FOD of the 
logarithm of the signal \cite{Klapuri:99}. 

\begin{equation}
RDF(t) = {1 \over \Delta t}(log(S(t)) - log(S(t-1)))
\label{RDF}
\end{equation}

\begin{figure}[thp]
  \begin{center}
    \resizebox{5in}{!}{\includegraphics{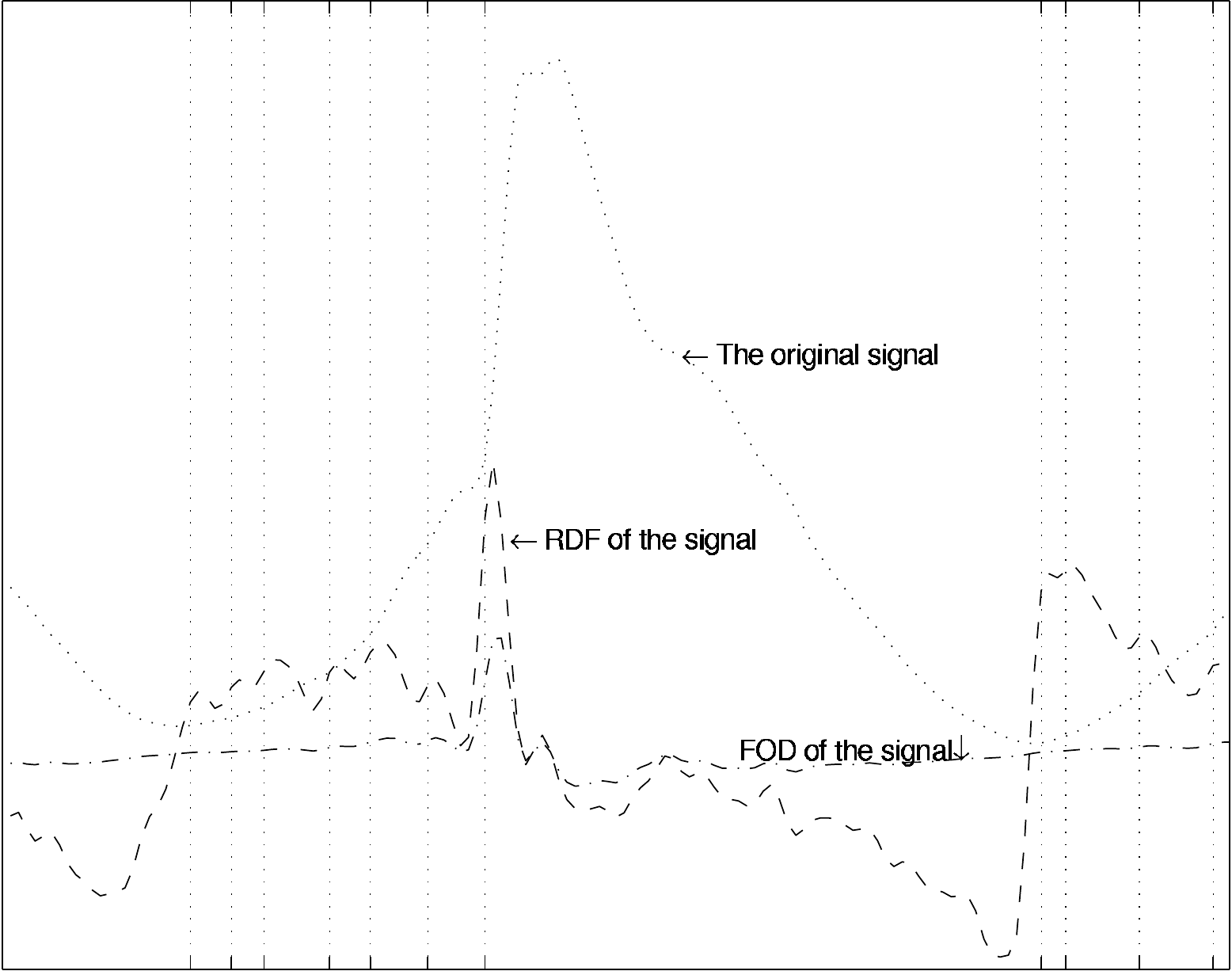}}
    \label{Comparison of First Order Difference and Relative
      Difference Functions}
    \caption{Comparison of the First Order Difference and Relative
      Difference function of a signal, the grid lines correspond to
      peaks in the RDF above a particular threshold}
  \end{center}
\end{figure}

The RDF solves the above mentioned problems since its peaks 
correspond to the beginning of the physical onset, thereby
resolving the problem of late onset estimation with gradually
increasing sounds. Moreover, once a signal starts to rise, local
minima and maxima along the attack slope are not significant relative
to the signal's level.

\begin{figure}[thp]
  \begin{center}
    \resizebox{5in}{!}{\includegraphics{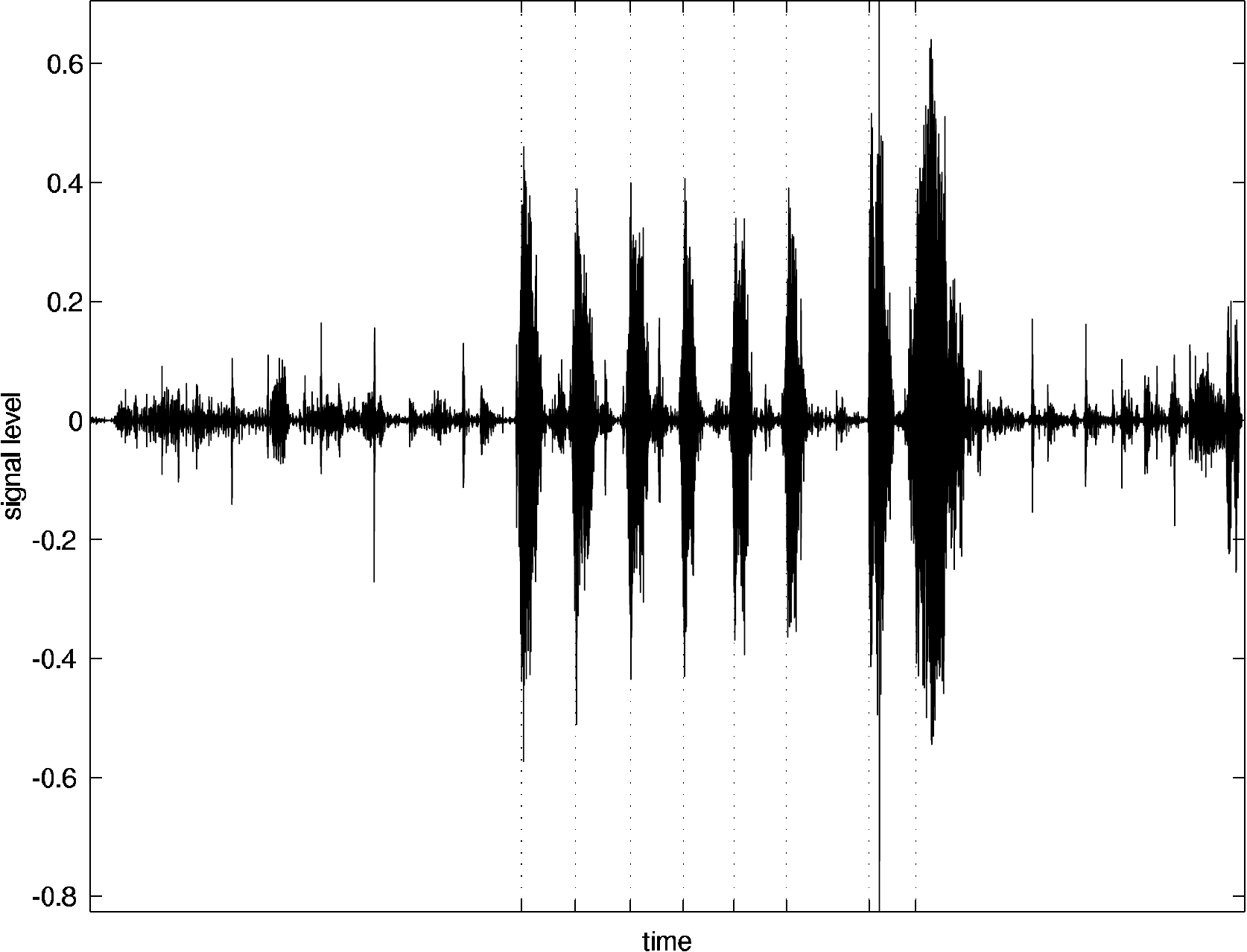}}
    \caption{A separated Soukous vocal chant, gridlines indicate time
      points where onsets are detected. The threshold value is set to .3}
    \label{vocalsampleonsets}
  \end{center}
\end{figure}

So the output of this module returns for each stream it processes, 
a sequence of time values in the stream when the onsets occur. If for
example, an extracted snare drum pattern was passed, 
it will return the start times of each snare hit. This information 
is significant for rhythmic analyses since the difference in start
times or inter-onset intervals (IOIs) are important in the
composition of rhythmic patterns. The snare hit ``off-times'' are less
important from a rhythmic point of view since once the onset occurs
and that attack has happened what follows is less important for the
determination of a rhythmic percept.

\begin{figure}[thp]
  \begin{center}
    \resizebox{4in}{!}{\includegraphics{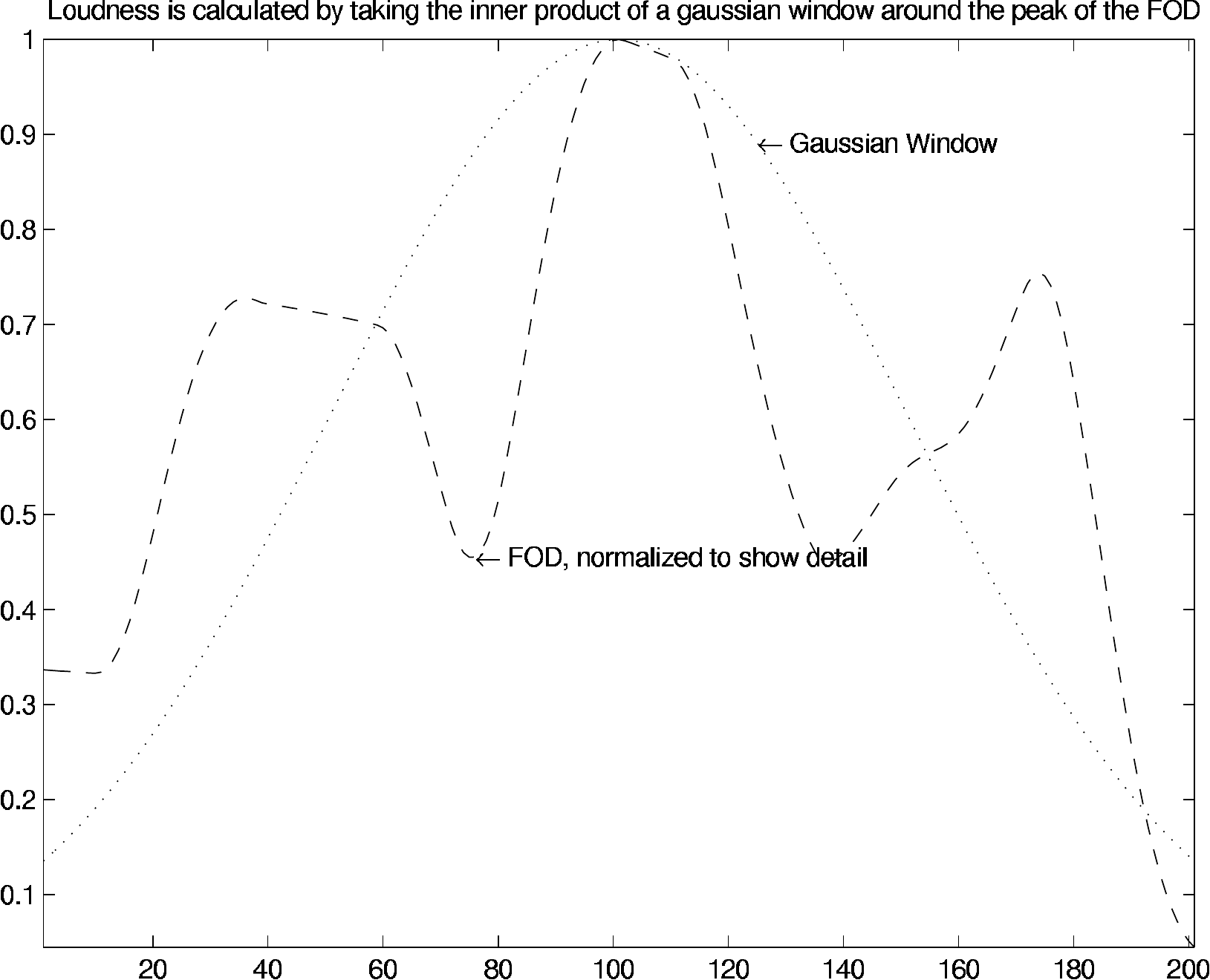}}
    \caption{The FOD represents well the loudness of an onsetting
      sound, the loudness is thus calculated by taking the inner
      product of a Gaussian window centered about the FOD peak in question}
    \label{Loudness}
  \end{center}
\end{figure}

During the determination of the peaks in the RDF, the loudness of each
peak is also calculated. The inner or dot product of the FOD
of the signal is taken with a Gaussian window equivalent in length to 
200 msec. The peak of the FOD coincides with the middle point 
of the normal Gaussian window. Refer to Figure \ref{Loudness}.

After the onsets are obtained as well as their respective sound pressure
levels, the timing data is run through a {\it pruning} routine. Pruning
helps eliminate spurious onsets due to imperfections present in the
ISA separation.  This routine takes the minimum spacing between onsets as
an argument set by the user. While the perceptual limit for
discriminable sounds tends to be around one sixteenth of a second,
percussive flams and grace note flourishes can be much more
rapid. Along with the spacing parameter for pruning, users can also set a
threshold value.  This is useful in dealing with anomalies
in the extracted signal that are usually at a much lower level than the salient
rhythmic features that characterize a stream.  Thus by setting an appropriately high
threshold and a minimum onset interval, clean timing and loudness data
can be gleaned from each stream.

\begin{figure}[thp]
  \begin{center}
    \resizebox{4in}{!}{\includegraphics{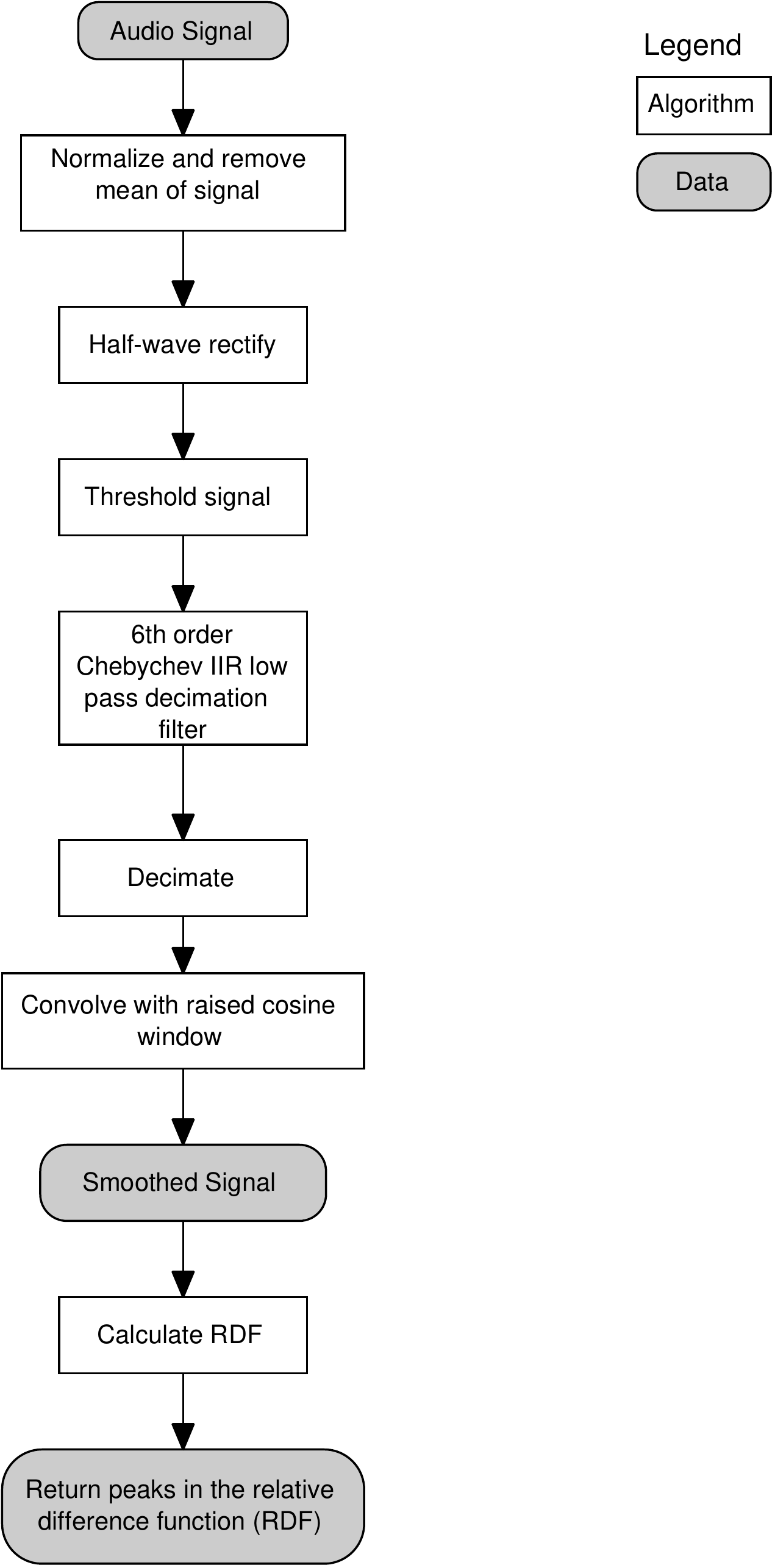}}
    \caption{Onset Detection Data Path}
    \label{Onset Detection}
  \end{center}
\end{figure}

%
% The first interpretation of the timing information ...
%
\vspace{7mm}
\section{Determination Of A Per-Stream Lowest Level Pulse}
\vspace{3mm}

% Furthermore any kind of musical information is open to multiple interpretation, 
% automated or otherwise.  With this in mind the ``interpretation'' of the timing
% information in terms will have a main candidate for grid size and other
% auxiliary ones.  This aspect in many previous computational models has been
% ignored because ambiguity is ambiguously represented [cite Desain Can Comp Models]

Having successfully separated individual streams from a sound
mixture, each potentially corresponding to a different instrument or 
sound source, timing information is then extracted. So far, this is 
the first of the two high level functions discussed above. 

The next step is to make rhythmic sense of the timing information.  
For example, with the per-stream time vectors we could try to find 
hierarchical time or grouping structures. We can also try to estimate
the overall and per-stream pulse or if applicable 
estimate quantities like meter and strength of beats. Due to time 
constraints and the limited scope of this thesis, I implemented only 
one interpretive module, one that estimates the lowest level pulse per-stream.

This idea of each stream having its own pulse is a deviation from all
previous work in the field which has tried, given a piece of music, to
estimate the {\it overall} lowest level pulse \cite{Scheirer:98tempo}.
If, however, pulse information is clearly determined for {\it each} 
instrument in an ensemble, then it is possible to examine a variety 
of sophisticated rhythmic inter-relationships between instruments.

\vspace{5mm}
\subsection{The Greatest Common Divisor (GCD) of the IOIs}

The algorithm I used to determine the per-stream pulse was
incidentally also from Tampere University of Technology (TUT), 
this time by Jarno Sepp\"anen. It is based on finding the 
``temporally shortest and perceptually lowest-level pulse''
\cite{Seppanen:99}. This idea of a temporal atom is not new 
and has been suggested by others as tatum, quantum or clock \cite{Seppanen:99}\cite{Bilmes:93}. 
Estimating the tatum of a piece of music is equivalent to approximating the 
greatest common divisor (GCD) of the all the inter-onset intervals (IOI) 
present in the music. 

First some definitions, the GCD of a set of natural numbers $a_i, i
\in I$ is defined to be the largest integer which divides all $a_i$ with a
zero remainder: 
\begin{equation}
gcd(a_1,a_2,...a_n) = max \{d: \forall i \in I: a_i \bmod r = 0 \}
\label{gcd}
\end{equation}
There is a problem however. The inter-onset intervals that are
calculated are not always integral. This is because we are
dealing with real performance data, there are never any such
guarantees. Sepp\"anen suggests that we try to find a tatum that
adequately divides all the IOIs and then define an error function
whose local minima represent the best candidates for the GCD. The
error function proposed is defined as such
\begin{equation}
e(q) =  \sum_{i}[(o_i + q/2) \bmod q - q/2]^2
\label{errorfunc}
\end{equation}
It is loosely based on the error function of the form
$\sum_{i}(o_{i}\bmod q)^2$. So if a GCD exists then the remainder
function will be zero at the GCD, otherwise the best approximation of
the GCD will be the the greatest value $q$ where $e(q)$ has a local minimum.
An assumption made by the remainder function is that IOI deviations
are gaussian in nature. Barring expressive timing modulations,
Scheirer has shown experimental evidence that timing deviations tend to be
distributed normally \cite{Scheirer:98}.

\vspace{5mm}
\subsection{Histogram Representation of Intra Frame IOIs}

Sepp\"anen's algorithm is designed to run in real time, processing data in
frames of 500 msec using a leaky integrator that ``remembers'', with
some decay, data from previous frames.  My implementation of this algorithm
is similar despite its non-realtime operation, because the net effect of listening
to music in realtime to determine a pulse is identical to analyzing a
recording causally.

Because 500 msec analysis frames are a bit short to capture rhythmic modulations, 
a histogram data structure is used to store IOI data from frame to frame.
Each time the histogram data structure is updated with IOI information from
the current frame, the newest additions are weighted higher while
histogram data from previous frames are weighted lower with some decay value. 

As IOI data is accumulated, it is first discretised to be a multiple of
the reciprocal of the sample rate $f_{s}$.  If $f_{s} = 44100$ there will
be $44100$ bins in the histogram each $1/44100$ seconds wide. This also means
that the upper bound on the lowest pulse is one second, as an IOI larger than
one second will ``fall off'' the histogram. This is a valid constraint as it
matches psychological experiments where events that are spaced more than 
two seconds apart are not able to be understood in a rhythmic context \cite{Warren:93}.

As we accumulate IOI data in the histogram form, each bin of the histogram
with a value greater than zero can be seen as an element of a set for which
we must find the GCD.  The error function defined above now must be expressed
in terms of the histogram bin values. Sepp\"anen defines such a function
as $$e(q) ={ \sum_{k=0}^{M-1}h[k][(h_{x} + q/2)\bmod q - q/2]^2 \over
  \sum_{k=0}^{M-1}h[k]}$$ where $h[k]$ is a histogram bin value and
$e(q)$ is normalized by the instantaneous histogram mass $\sum_{k=0}^{M-1}h[k]$. Once the error function is calculated,
the largest indice containing a local minimum is the candidate for
GCD.  Picking this candidate involves a traversal of the error
function, calculating first and second order difference functions and 
finding indicies in the histogram that correspond to peaks in the FOD 
that that have positive SOD values (i.e. concave up)

\begin{figure}[thp]
  \begin{center}
    \resizebox{4in}{!}{\includegraphics{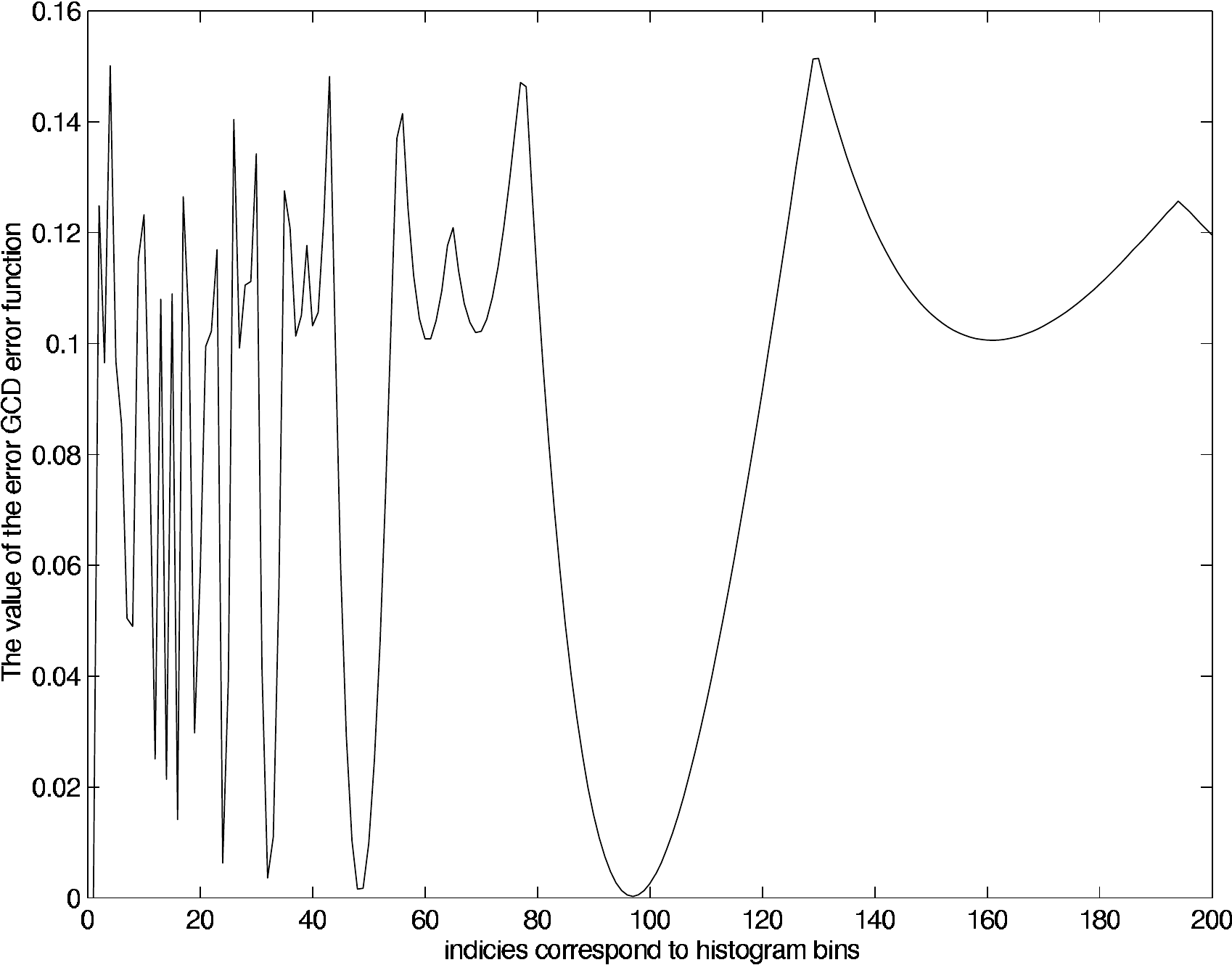}}
    \caption{The error function from one 500 msec frame of onset
      times extracted from the soukous kick drum. The largest indice
      with the lowest local minima is 96. Thus the lowest level pulse
      corresponds to the time in seconds represented by histogram bin 96} 
    \label{The GCD Error Function $e(q)$}
 \end{center}
\end{figure}

As the signal's IOI values are successfully accumulated in the
histogram, the arrival of each new frame triggers an update of the error
function and the estimation of the histogram bin containing the GCD. Accordingly,
over the course of a ten second piece of music as many as 20 estimations will
be made for the lowest level pulse, thus capturing rhythmic feature modulations.
This per-instrument low level pulse trajectory is then plotted over
incrementally for the duration of the musical segment in question.

\begin{figure}[thp]
  \begin{center}
    \resizebox{5.2in}{!}{\includegraphics{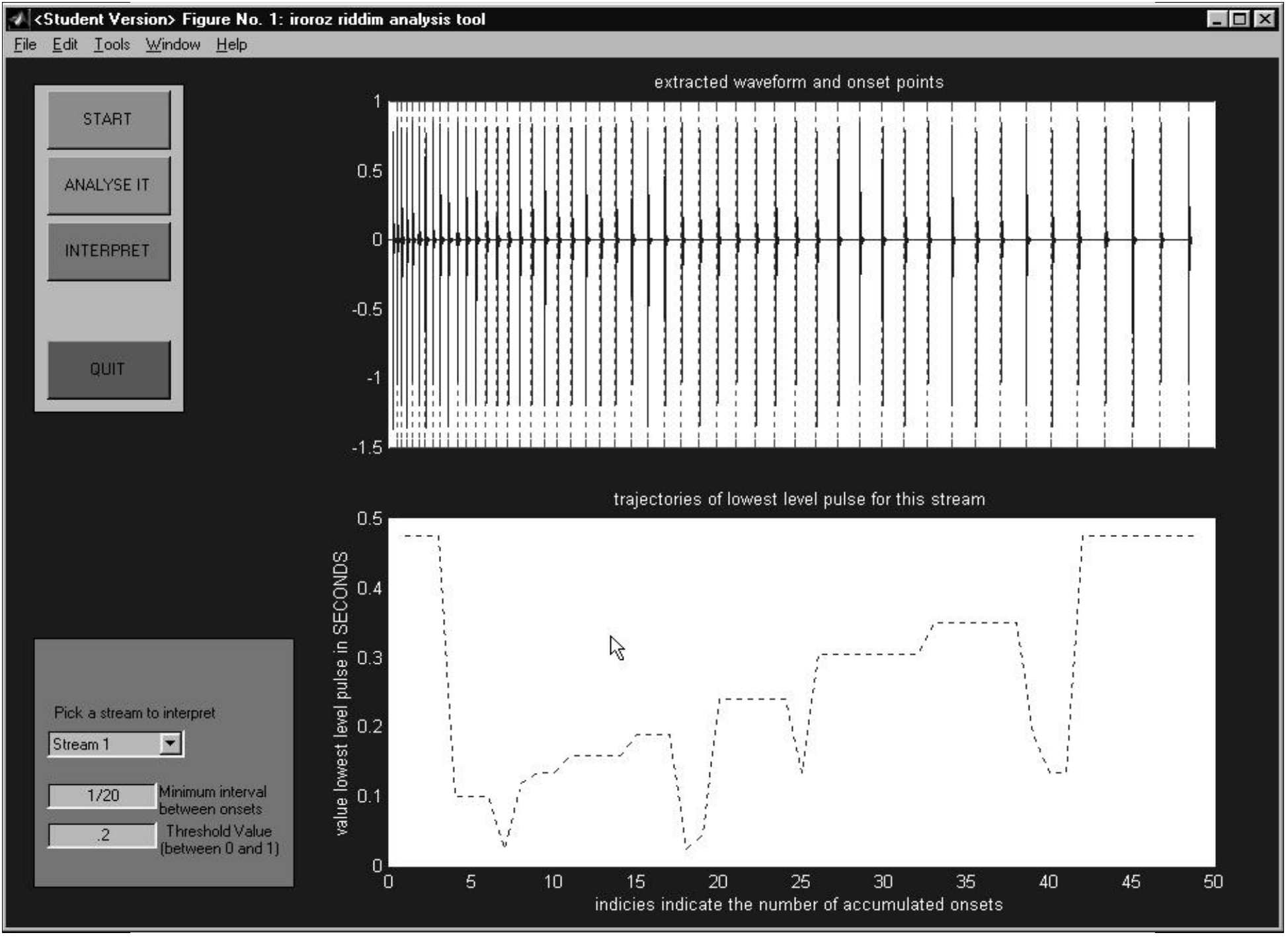}}
    \caption{A plot of the trajectory of the lowest level pulse}
    \label{lowestlevelpulse}
  \end{center}
\end{figure}

  \ifnum0=\value{mychaptercount}
    \startingpages
    \setcounter{mychaptercount}{1}
  \fi
  \chapter{Results}
\vspace{10mm}

\begin{quote}
  {\it ``There's a basic rule which runs through all kinds of music, 
    kind of an unwritten rule. I don't know what it is.  But I've got
    it.''} --- Ron Wood, guitarist for the Rolling Stones.
\end{quote}

\vspace{7mm}

The result of this thesis is a software system that 
realizes the algorithms described in the previous chapter. 
The outputs of the software are twofold. First, there is the rhythmic 
timing data extracted from each stream that is saved as a MIDI file or
re-interpreted  with an audio sample and saved as digital audio.
Second, there is the graphical interpretation of the timing data as the 
lowest-level pulse in each stream.

% [Discuss here the process of mixing 3 signals and then running them
% thru ISA and talk about the results]

The quality of the MIDI rendering and the re-interpreted audio files is
very promising.  In most cases, a high quality extraction of an
instrument virtually guarantees clean results from the onset
analysis. Additionally, variable threshold values and minimum
onset spacing parameters add to the robustness of the onset detection
algorithm. The ISA algorithm works especially well when the number of
voices to be separated is low (i.e. less than five). Note that when the
number of voices is much higher, a hierarchical ISA technique 
should be employed. This means that a mono audio file is first run through a
ISA extraction with a low number of components, and each component is
then run through subsequent ISA extractions to further unmix the
latent sources.

The interpretation of the timing data for each stream is also promising.
While not having many practical compositional uses, its
analytic value is important.  For example, an accelerando or
ritardando in an extracted instrument will be seen as an increase or 
decrease respectively in that instrument's lowest level
pulse. In most popular dance and folk music, if one of the extracted 
instruments is a kick drum or some other salient pulse keeper, the
lowest  level pulse is interpreted as the foot-tapping beat or tactus 
of the entire work.

\begin{figure}[thp]
  \begin{center}
    \resizebox{4.5in}{!}{\includegraphics{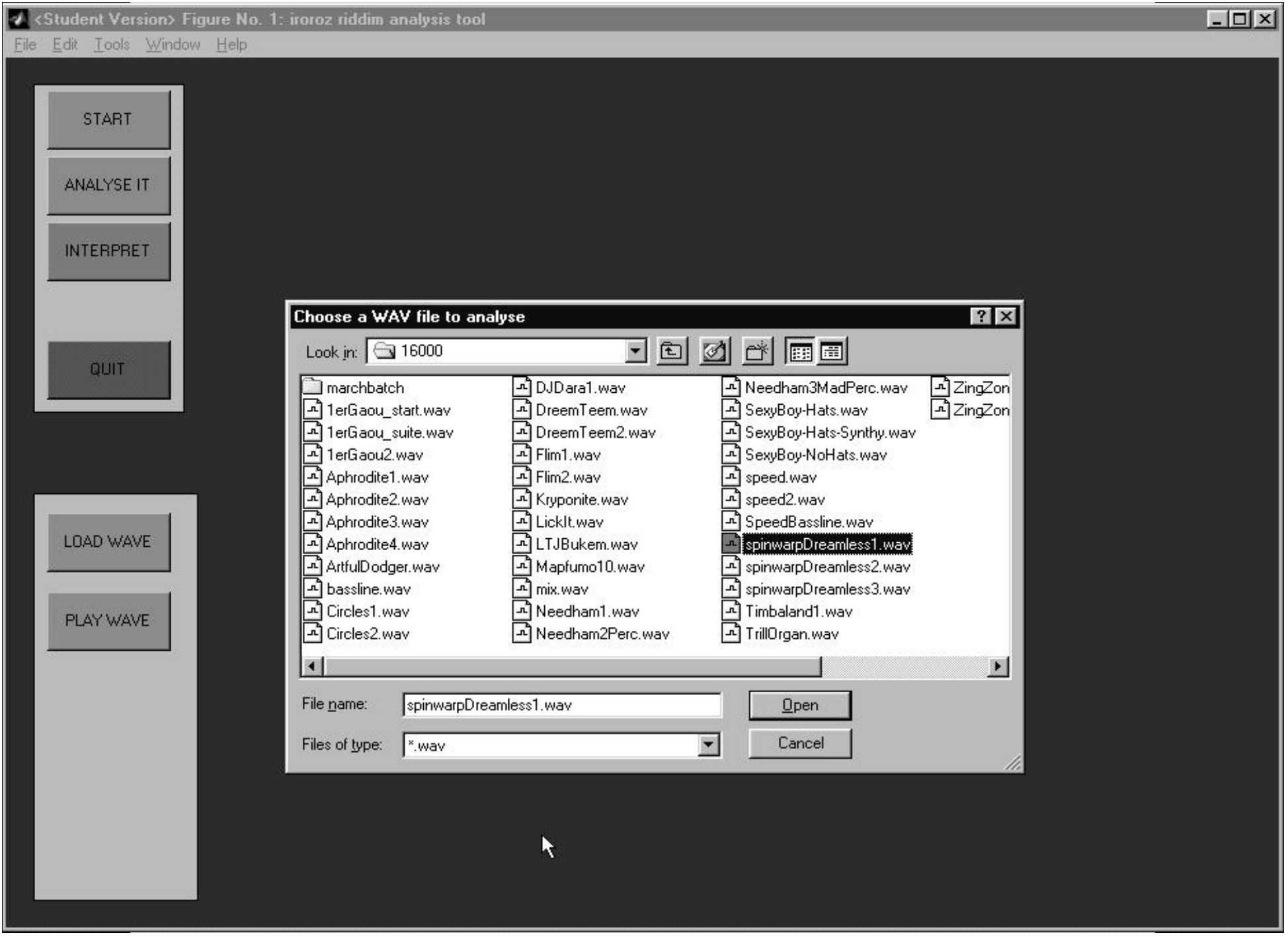}}
    \caption{A screen shot of {\it Riddim} in the Start Mode, with an
      open dialog box to choose a WAV file to analyse}  
    \label{Riddim Analysis Start Screen Shot}
  \end{center}
\end{figure}

\vspace{7mm}
\section{{\it Riddim}}
\vspace{3mm}

Because {\it Riddim} is also a proof-of-concept for a real-time rhythm
analysis/synthesis engine, the completion of the software goes beyond
the scope of this thesis. For that reason, I will not be including 
the source code in this document. Rather, all the source and examples 
will be distributed on the web under the GNU General Public License
(GPL) agreement, available at \begin{verbatim}
  http://eamusic.dartmouth.edu/~iroro/
\end{verbatim}

{\it Riddim} is organized into three modes, a start, an analysis 
and an interpretation mode. In the start mode, the user can load and
preview audio files.  The user then advances to the analysis mode   
where the loaded audio data is subjected to the algorithms 
discussed in the previous chapters.  The user can enter a set
of variables that direct the course of the analysis algorithms. After
the user pushes ``GO'', the ISA extraction routines proceed and return a
menu allowing the user to select a stream to view. 
By selecting a stream, the onset detection algorithm is run and the 
user is presented with the extracted waveform with superimposed
grid lines indicating onset times. At this stage the
user is can enter a spacing parameter and a threshold value to
fine tune the quality of the onset detection. The user change also
change the parameters used for the ISA extraction by hitting
``GO'' to re-run the analysis. \begin{figure}[thp]
  \begin{center}
    \resizebox{4.5in}{!}{\includegraphics{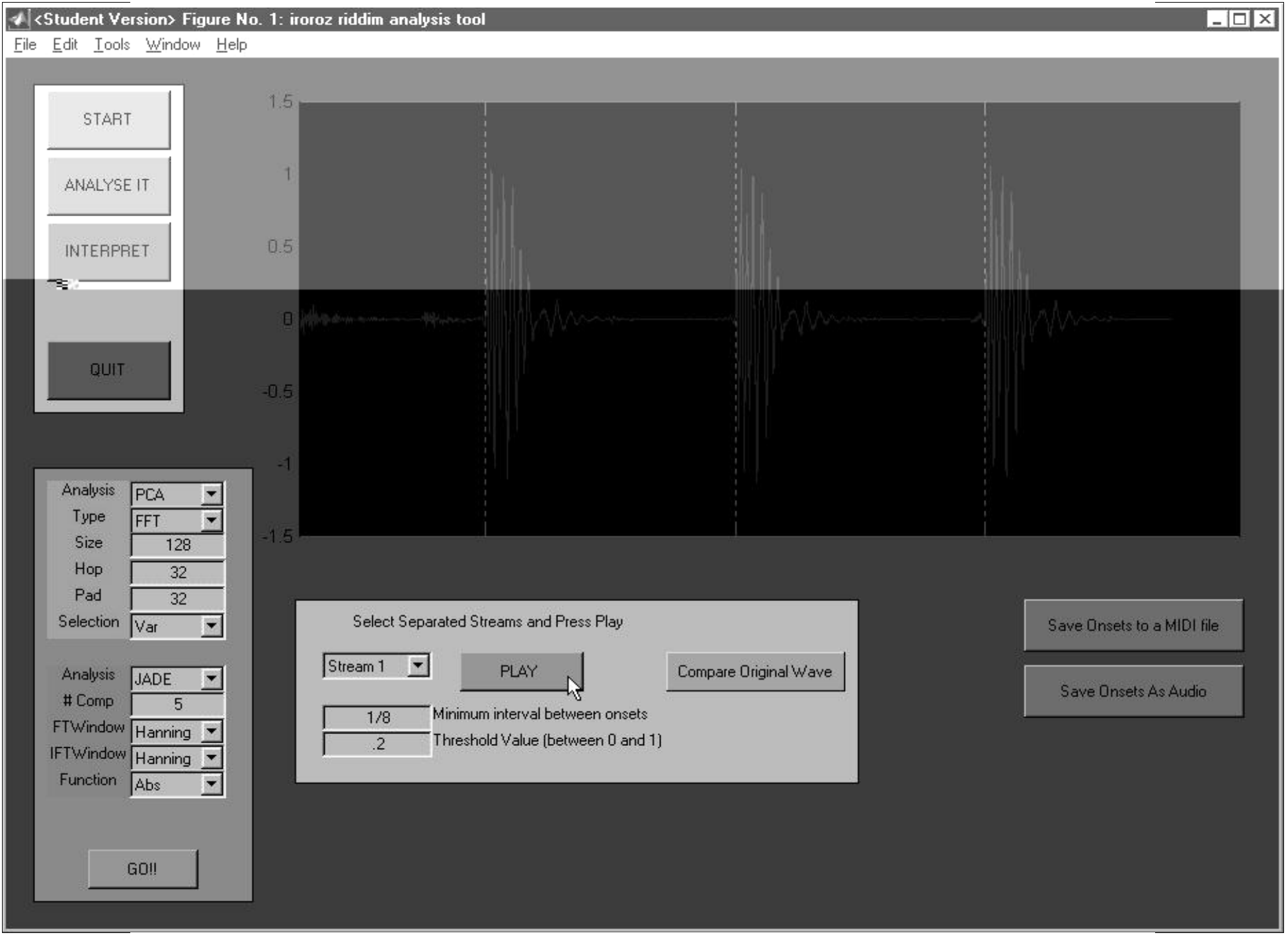}}
    \caption{A screen shot of {\it Riddim} in the Analysis Mode. Shown
      are an extracted kick drum and the detected onsets indicated by
      the grid lines.} 
    \label{Riddim Analysis Screen Shot}
  \end{center}
\end{figure}
Once the onset times per stream are 
as desired, there are two buttons that allow the timing
information to be saved as a MIDI file or as digital audio.  In the
latter case, a dialog window will appear asking the user to choose a 
file with which to render the timing information. At any stage the
user can load another audio file or reload the current file.

So far these modes embody the functionality of the first high level
subsystem. The next and final mode, the ``Interpretation'' mode, is the
forum for a variety of interpretive analyses of the extracted timing
data. As was discussed above, only the determination of the
lowest-level pulse was implemented. Accordingly, in this mode, the
user is presented with the same menu of extracted streams. This time 
selecting a stream will return the specified stream, the associated 
onsets {\sl and} a plot of the movement of the lowest level
pulse. Refer to Figure \ref{lowestlevelpulse}.

  \ifnum0=\value{mychaptercount}
    \startingpages
    \setcounter{mychaptercount}{1}
  \fi
  \chapter{Conclusions and and Future Directions} 
\vspace{10mm}

\begin{quote}
  {\it ``I conclude that musical notes and rhythms were first acquired
    by the male and female progenitors of mankind for the sake of
    charming the opposite sex.''} --- Charles Darwin
\end{quote}

\vspace{7mm}
\section{Automatic Music Transcription}
\vspace{5mm}

The most immediate application of the ideas behind {\it Riddim}
is in an automatic music transcription system.  Traditionally, 
transcription involves writing down the notes occurring in a
piece of music, converting an acoustic signal to a symbolic
representation \cite{Klapuri:98}. Learning to 
transcribe music involves however, a lifetime of musical training
and specific training in the styles to be transcribed. An
automatic transcription software tool thus saves time 
and is useful for musicians and composers for a variety of 
analytic and pedagogical purposes. For example, in many modern styles 
where a score is not available, a symbolic representation of what is 
occurring in the music is very valuable. Similarly, Jazz buffs trying 
to learn the intricacies of an improvised jazz solo can benefit from 
such a system as well as can academic composers trying to analyse an 
electro-acoustic work.

To make {\it Riddim} a robust music transcription system will 
entail additional work. First, a fundamental frequency 
tracking module is essential for a classic ``score'' transcriber. 
This module would essentially traverse the extracted streams 
performing a pitch estimation at every onset point. In this
way, the onset times would correspond to the rhythmic part
of the score while the estimated fundamental frequency would
approximate the pitch. Invariably, as audio analysis techniques become more sophisticated,
automatic music transcription systems like speech recognition
systems are going to be increasingly important in the way people
learn and interact with music. As Klapuri points out, ``Some people
would certainly appreciate a radio receiver capable of tracking
jazz music simultaneously on all channels'' \cite{Klapuri:98}.

\vspace{7mm}
\section{{\it Riddim} in Real-time}
\vspace{5mm}

One of the central motivations of this thesis was to take the music
that I have composed from the studio to the stage, without compromising 
the complexity. At the same time, moving to a more improvisational 
performance setting requires a certain mastery of the musical content 
and the tools to be able to compose on the fly. The current work is of
course non-realtime. Working with the algorithms and the code for the
past nine months, I have determined that a realtime version is not
only feasible, but with minimal additional work, can run as a
MAX/MSP{\texttrademark} external DSP module or as a VST{\texttrademark} plugin.

\vspace{7mm}
\section{Live Improvised Performance and the Meta DJ}
\vspace{5mm}

Once running in real-time, what are its musical applications? 
In a live performance setting where listening and playing within 
a shared musical context is paramount, {\it Riddim} ``listens'' to the surrounding
music, analyses it and reveals patterns that can be used time a new
instrument or part. How these patterns are mixed with the original
material is a delicate question of production and aesthetics, as the
choice of voicing makes a world of difference.

Furthermore, in analysing and re-interpreting music on the fly, mixing
elements occurs in an abstract representation of the music rather than the 
common DJ practise of mixing the concrete representation of the music i.e. samples.
The idea of a {\it Meta DJ} involves mixing an abstract representation
of ``vinyl cuts'' or CD tracks by synchronising the realtime outputs of {\it Riddim} and
rendering the patterns with variety of instruments appropriate for the
musical setting. The flexibility of not having to mix concrete samples
is significant since resampling, time-stretching and vocoder artifacts are common
byproducts trying to match two tracks at different tempi. Furthermore, the time patterns
extracted in realtime can be used to control light shows, smoke
machines or any number of patterned sensory stimuli that accompany the
music.

%\newchapter{chapter6}

% Include any appendices.
%\newappendix{appendixA}
%\newappendix{appendixB}

\end{document}